\documentclass[11pt]{article}
\pdfoutput=1
\usepackage{amsfonts,amssymb,amsmath,amscd,amsthm,latexsym}
\usepackage{natbib}
\usepackage{graphicx}

\newcommand{\ds}{\displaystyle}

\usepackage[normalem]{ulem}

\newcommand{\bx}{\mathbf{x}}
\newcommand{\bX}{\mathbf{X}}
\newcommand{\by}{\mathbf{y}}
\newcommand{\knn}{$k$-nearest-neighbour}

\title{A Bayesian reassessment of nearest--neighbour classification}

\author{Lionel Cucala${}^{1}$, Jean-Michel Marin${}^{1,3}$,
Christian P.~Robert${}^{2,3}$ \\ and D.M. Titterington${}^4$\\
${}^1$INRIA Saclay, Projet \textsc{select}, Universit\'e Paris-Sud,\\
${}^2$CEREMADE, Universit\'e Paris Dauphine,\\
${}^3$CREST--INSEE, and ${}^4$University of Glasgow}

\begin{document}

\maketitle
\begin{abstract}
The $k$-nearest-neighbour procedure is a well-known deterministic method used in
supervised classification. This paper proposes a
reassessment of this approach as a statistical technique derived from a proper probabilistic 
model; in particular, we modify the assessment made in a previous analysis of this method 
undertaken by \cite{Holmes:Adams:2002,Holmes:Adams:2003}, and evaluated by \cite{Manocha:Girolami:2007},
where the underlying probabilistic model is not 
completely well-defined. Once a clear probabilistic basis for the $k$-nearest-neighbour procedure is 
established, we derive computational tools for conducting Bayesian inference 
on the parameters of the corresponding model. In particular, we assess the difficulties inherent to 
pseudo-likelihood and to path sampling approximations of an intractable normalising constant, and propose
a perfect sampling strategy to implement a correct MCMC sampler associated with our model.
If perfect sampling is not available, we suggest using a Gibbs sampling approximation.
Illustrations of the performance of the corresponding Bayesian classifier are provided for 
several benchmark datasets, demonstrating in particular the limitations of the pseudo-likelihood approximation
in this set-up.\\

\end{abstract}

\section{Introduction}

\subsection{Deterministic versus statistical classification}

Supervised classification has long been used in both Machine Learning and 
Statistics to infer about the functional connection between a group of covariates (or 
explanatory variables) and a vector of indicators (or classes)
\citep[see, e.g.,][]{McLachlan:1992,Ripley:1994,
Ripley:1996,Devroye:Laszlo:Lugosi:1996,Hastie:Tibshirani:Friedman:2001}.
For instance, the method of {\em boosting} \citep{Freund:Schapire:1997} has been developed for this very 
purpose by the Machine Learning community and has also been assessed and extended by statisticians 
\citep{Hastie:Tibshirani:Friedman:2001,Buhlmann:Yu:2002,Buhlmann:Yu:2003,Buhlmann:2004,Zhang:Yu:2005}.

The \knn~method is a well-established and straightforward technique in this area with both a long past and
a fairly resilient resistance to change \citep{Ripley:1994,Ripley:1996}. Nonetheless,
while providing an instrument for classifying points into two or 
more classes, it lacks a corresponding assessment of its classification error. While alternative techniques 
like boosting offer this assessment, it is obviously of interest to provide the original \knn~method with 
this additional feature. In contrast, statistical classification methods that are based on a model such a 
mixture of distributions do provide an assessment of error along with the most likely classification.
This more global perspective thus requires the technique to be embedded within a probabilistic framework in order to
give a proper meaning to the notion of classification error. \cite{Holmes:Adams:2002}
propose a Bayesian analysis of the \knn-method based on these premises, and we refer the reader to this 
paper for background and references. 
In a separate paper, \cite{Holmes:Adams:2003} defined another model based on
autologistic representations and conducted a likelihood analysis of this model, in particular for selecting
the value of $k$. While we also adopt a Bayesian approach, our paper differs from 
\cite{Holmes:Adams:2002} in two important respects: first, we define a global probabilistic model that
encapsulates the \knn~method, rather than working with incompatible conditional distributions, and, second, 
we derive a fully operational simulation technique adapted to our model and based either on perfect sampling
or on a Gibbs sampling approximation, that allows for a reassessment of the pseudo-likelihood approximation 
often used in those settings.

\subsection{The original \knn~method}\label{sec:origami}
Given a training set of individuals allocated each to one of $G$ classes,
the classical $k$-nearest-neighbour procedure is a method that allocates new individuals to 
the most common class in their neighbourhood among the training set, the neighbourhood being defined 
in terms of the covariates.  More formally, 
based on a training dataset $\left((y_i,x_i)\right)_{i=1}^n,$ where $y_i\in\{1,\ldots,G\}$ denotes 
the class label of the $i$th point and $x_i\in\mathbb{R}^p$ is a vector of covariates, an 
unobserved class $y_{n+1}$ associated with a new set of covariates $x_{n+1}$ is estimated by 
the most common class among the $k$ nearest neighbours of $x_{n+1}$ in the training set 
$\left(x_i\right)_{i=1}^n$. The neighbourhood is defined
in the space of the covariates $x_i$, namely
$$
\mathcal{N}^k_{n+1} = \left\{ 1\le i\le n;\,d(x_i,x_{n+1}) \le d(\cdot,x_{n+1})_{(k)}\right\}\,,
$$
where $d(\cdot,x_{n+1})$ denotes the vector of distances to $x_{n+1}$ and $d(\cdot,x_{n+1})_{(k)}$
denotes the $k$th order statistic. The original \knn~method usually uses the Euclidean norm, 
even though the Mahalanobis distance would be more appropriate in that it rescales the covariates.
Whenever ties occur, they are resolved by decreasing the number $k$ of neighbours until the
problem disappears. 
When some covariates are categorical, other types of distance can be used instead, as in the 
{\sf R} package {\sf knncat} of \cite{buttrey:1998}.

As such, and as also noted in \cite{Holmes:Adams:2002}, the method is both deterministic, 
given the training dataset, and not parameterised, even though the choice of $k$ is 
both non-trivial and relevant to the performance of the method. Usually, $k$ is 
selected via cross-validation, as the number of neighbours that minimises the
cross-validation error rate. 
In contrast to cluster-analysis set-ups, the number $G$ of classes in the $k$-nearest-neighbour procedure
is fixed and given by the training set: the introduction of additional classes
that are not observed in the training set has no effect on the future allocations.

\begin{figure}[hbtp]
\centerline{
\includegraphics[width=180pt]{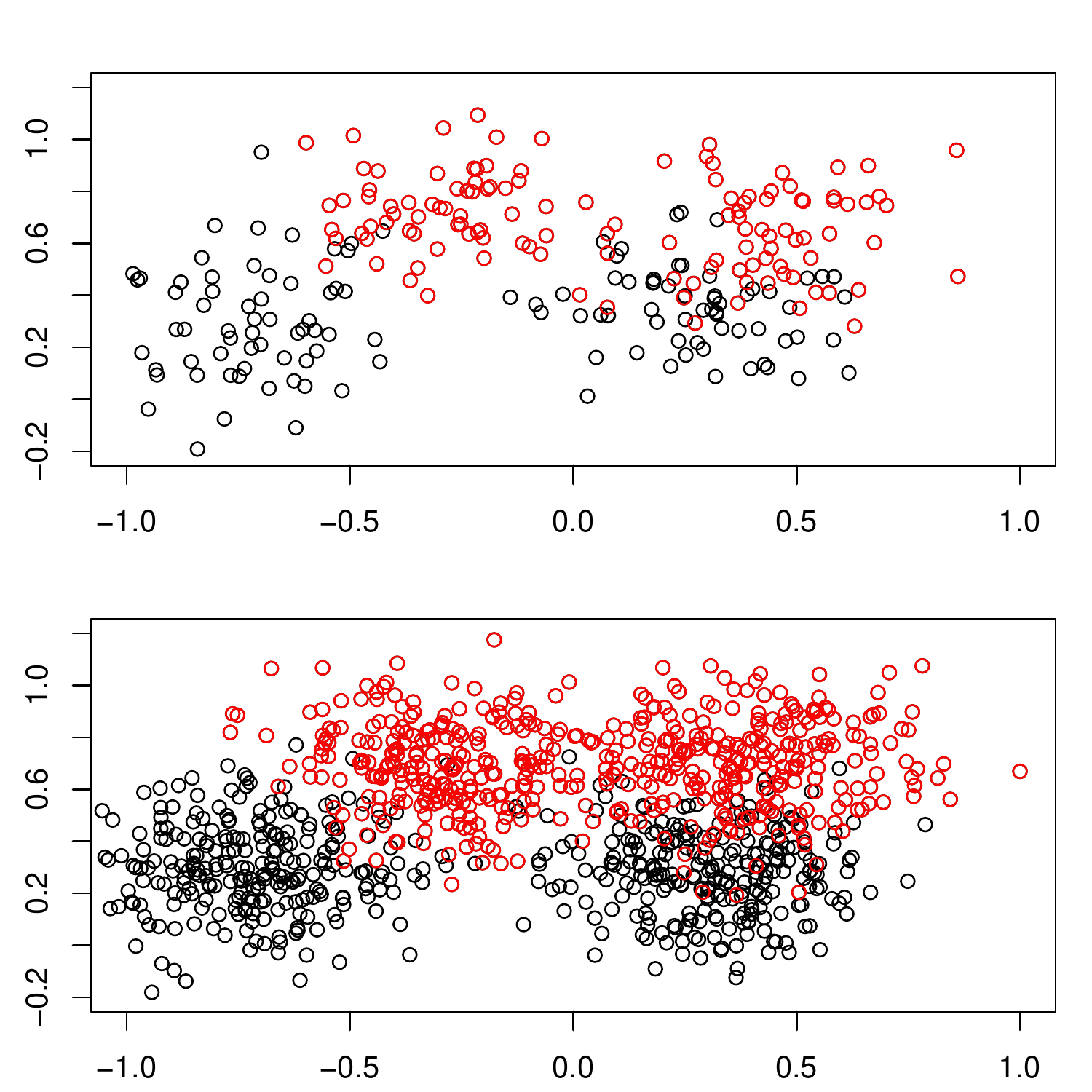}
}
\caption{\label{fig:ripley} Training {\em (top)} and test {\em (bottom)} groups for Ripley's 
benchmark: the points in red are those for which the label is equal to $1$ and the points in black
are those for which the label is equal to $2$.}
\end{figure}

\begin{figure}[hbtp]
\centerline{
\includegraphics[width=180pt]{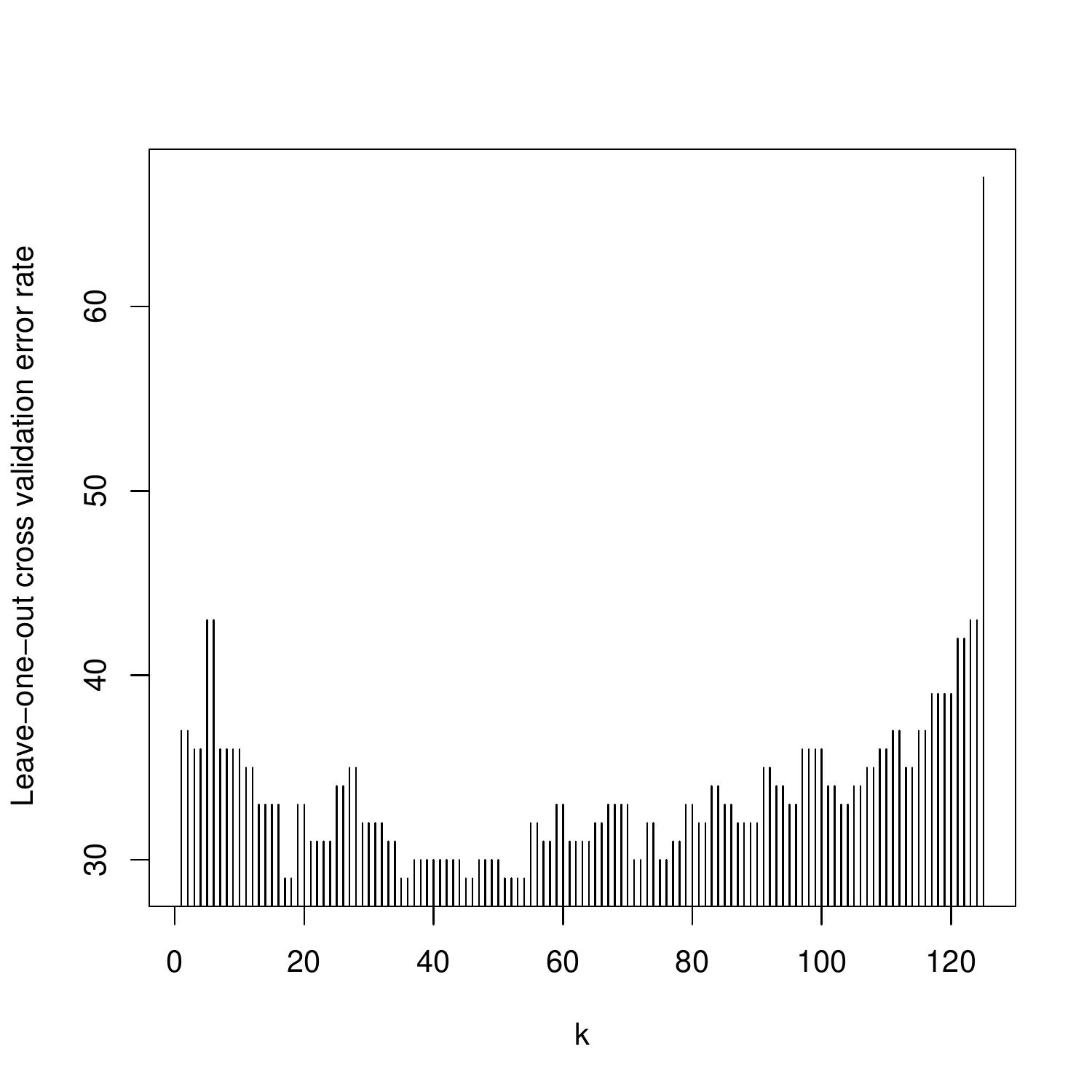}
}
\caption{\label{fig:leave1} 
Misclassification leave-one-out error rate as a function of $k$ for Ripley's training dataset.
}
\end{figure}

To illustrate the original method and to compare it later with our approach, we use throughout a toy benchmark dataset
taken from \cite{Ripley:1994}.  This dataset corresponds to a two-class classification problem 
in which each 
(sub)population of covariates is simulated from a bivariate normal distribution, both populations being of 
equal sizes. A sample of $n=250$ individuals is used as the training set and the model is tested on a 
second group of $m=1,000$ points acting as a test dataset. Figure \ref{fig:ripley} presents the 
dataset\footnote{This dataset is available at {\sf http://www.stats.ox.ac.uk/pub/PRNN}.} and
Table \ref{tab:knn-ripley} displays the performance of the standard \knn~method on the test dataset 
for several values of $k$. 
The overall misclassification leave-one-out error rate on the training dataset as $k$ varies is provided in Figure \ref{fig:leave1} and it shows
that this criterion is not very discriminating for this dataset, with little variation for a wide range of values of $k$ and with
several values of $k$ achieving the same overall minimum, namely
$17$, $18$, $35$, $36$, $45$, $46$, $51$, $52$, $53$ and $54$.
There are therefore ten different values of $k$ in competition. This range of values is an  indicator of potential
gains when averaging over $k$, and hence calls for a Bayesian perspective. 

\begin{table}[hbtp]
\begin{center}
\begin{tabular}{cc}
 $k$  & Misclassification \\
      & error rate \\
\hline
 1    & 0.150 \\    
 3    & 0.134 \\ 
 15   & 0.095 \\ 
 17   & 0.087 \\
 31   & 0.084 \\
 54   & 0.081 \\
\hline
\end{tabular}
\end{center}
\caption{\label{tab:knn-ripley} \knn~performances on the Ripley test dataset}
\end{table}

\subsection{Goal and plan}
As presented above, the \knn~method is merely an allocation technique that does not 
account for uncertainty. In order to add this feature, we need to introduce 
a probabilistic framework that relates the class label $y_i$ to both the covariates $x_i$ {\em and} 
the class labels of the neighbours of $x_i$. 
Not only does this perspective provide more information about the variability of the classification, 
when compared with the point estimate given by the original method, but it also takes advantage of 
the full (Bayesian) inferential machinery to introduce
parameters that measure the strength of the influence of the neighbours, and to analyse 
the role of the variables, of the metric used, of the number $k$ of neighbours, and of the number of classes 
towards achieving higher efficiency.  Once again, this statistical viewpoint was previously 
adopted by \cite{Holmes:Adams:2002,Holmes:Adams:2003} and we follow suit in this paper, with a modification of 
their original model geared towards a coherent probabilistic model, while providing new developments in computational 
model estimation.

\begin{figure}[hbtp]
\centerline{
\includegraphics[width=180pt]{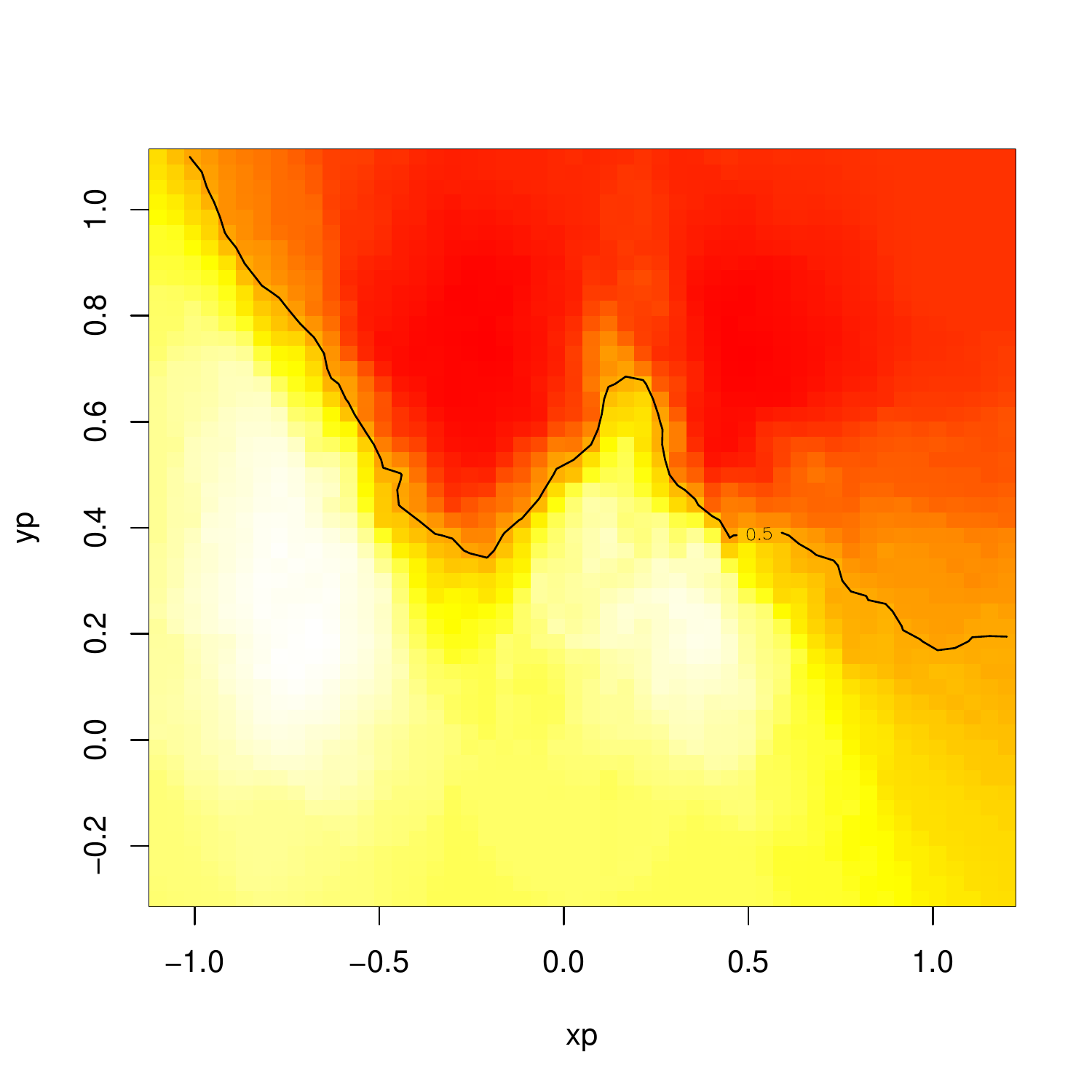}\includegraphics[width=180pt]{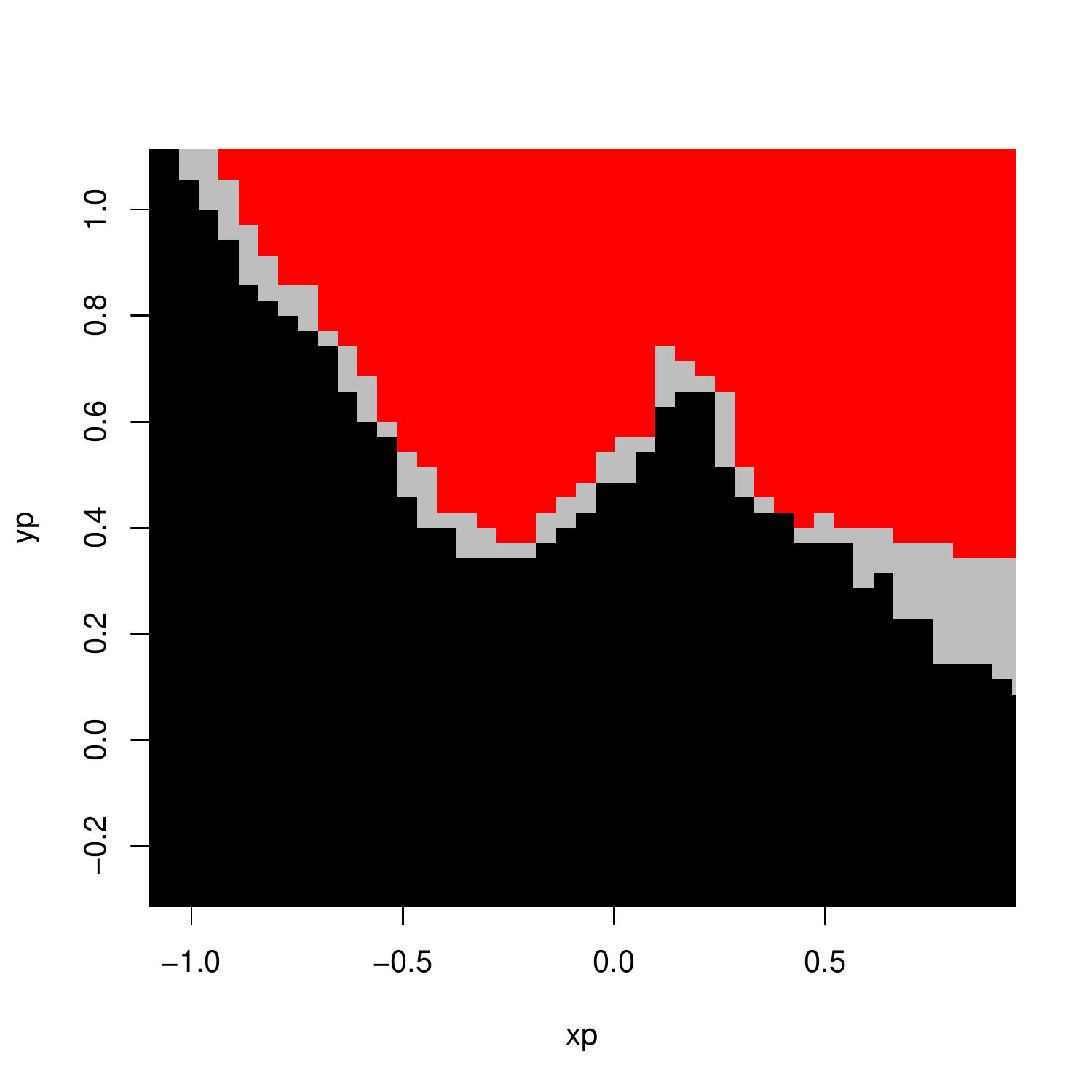}
}
\caption{\label{fig:levelset} 
{\em (left)} Level sets of the predictive probability to be in the black class, ranging from high {\em (white)}
to low {\em (red)}, and
{\em (right)} consequences of the comparison with $0.5$ of the $95$\%~credibility intervals for the 
predictive probabilities.
(These plots are based on an MCMC sample whose derivation is explained in Section \ref{sec:perfetto}.)
}
\end{figure}

In order to illustrate the appeal of adopting a probabilistic perspective, we provide in Figure \ref{fig:levelset} 
two graphs that are by-products of our Bayesian analysis.
For Ripley's dataset, the first graph (on the left) 
gives the level sets of the predictive probabilities to be in the 
black class, while the second graph (on the right) partitions
the square into three zones, namely sure allocation to the red class, sure allocation to the black class and an uncertainty zone.
Those three sets are obtained by first computing $95$\%~credible intervals for the predictive probabilities and then checking 
those intervals against the borderline value $0.5$. If the interval contains $0.5$, the point is ranked as uncertain.

The paper is organised as follows. We establish the validity of
the new probabilistic \knn~model in Section \ref{sec:model}, pointing out the deficiencies 
of the models advanced by \cite{Holmes:Adams:2002,Holmes:Adams:2003}, and then cover the different aspects of running
Bayesian inference in this \knn~model in Section \ref{sec:Binf}, addressing in particular the 
specific issue of evaluating the normalising constant of the probabilistic \knn~model that is 
necessary for inferring about $k$ and an additional parameter. We take advantage of an exact MCMC approach 
proposed in Section \ref{sec:perfetto} to evaluate the limitations of the pseudo-likelihood alternative in 
Section \ref{sec:psudo} and illustrate the method on several benchmark datasets in Section \ref{sec:illustre}.

\section{The probabilistic \knn~model}\label{sec:model}
\subsection{Markov random field modelling}
In order to build a probabilistic structure that reproduces the features of the
original \knn~procedure and then to estimate its unknown parameters, we first need 
to define a joint distribution of the labels $y_i$ conditional on the covariates $x_i$,
for the training dataset. A natural approach is
to take advantage of the spatial structure of the problem and to use a Markov random field 
model. Although we will show below that this is not possible within a coherent 
probabilistic setting, we could thus assume that the full conditional distribution of $y_i$ given
$\by_{-i}=\left(y_1,\ldots,y_{i-1},y_{i+1},\ldots,y_n\right)$ and the $x_i$'s
only depends on the $k$ nearest neighbours of $x_i$ in the training set. The parameterised structure of
this conditional distribution is obviously open but we opt for the most standard choice,
namely, like the Potts model, a Boltzmann distribution \citep{Moeller:Waagepetersen:2003} 
with potential function 
$$
\sum_{\ell\,\sim_k i} \delta_{y_i}(y_\ell)\,,
$$
where $\ell\,\sim_k i$ means that the summation is taken over the observations $x_\ell$
belonging to the $k$ nearest neighbours of $x_i$, and $\delta_a(b)$ denotes the Dirac function.
This function actually gives the number of points from the same class $y_i$ as the 
point $x_i$ that are among the $k$ nearest neighbours of $x_i$.
As in \cite{Holmes:Adams:2003}, the expression for the full conditional is thus
\begin{equation}
f(y_i|\by_{-i},\bX,\beta,k)=
{\ds \exp\left(\beta \sum_{\ell\,\sim_k i} \delta_{y_i}(y_\ell)\bigg/ k\right)}\Bigg/{\ds 
\sum_{g=1}^G \exp\left(\beta \sum_{\ell\,\sim_k i} \delta_{g}(y_\ell)\bigg/ k\right)}
\label{eq:knn-intuit}
\end{equation}
where $\beta>0$ and $\bX$ is the $(p,n)$ matrix $\{x_1,\ldots,x_n\}$ of coordinates
for the training set.

In this parameterised model, $\beta$ is a quantity that is obviously missing
from the original \knn~procedure. It is only relevant from a 
statistical point of view as a degree of uncertainty: $\beta=0$ corresponds to a 
uniform distribution over all classes, meaning independence from the neighbours,
while $\beta=+\infty$ leads to a point mass distribution 
at the prevalent class, corresponding to extreme dependence.
The introduction of the scale parameter $k$ in the denominator is useful in making $\beta$ dimensionless.

There is, however, a difficulty with this expression in that, for almost all datasets $\bX$,
there does not exist a joint probability distribution on $\by = (y_1,\ldots,y_n)$
with full conditionals equal to \eqref{eq:knn-intuit}. This happens because
the \knn~system is usually
asymmetric: when $x_i$ is one of the $k$ nearest neighbours of $x_j$, $x_j$ is not
necessarily one of the $k$ nearest neighbours of $x_i$. Therefore, the pseudo-conditional
distribution \eqref{eq:knn-intuit} will not depend on $x_j$ while the equivalent for $x_j$ does
depend on $x_i$: this is obviously impossible in a coherent probabilistic framework
\citep{Besag:1974,Cressie:1993} 

One way of overcoming this fundamental difficulty is to follow \cite{Holmes:Adams:2002} and to define 
directly the joint distribution
\begin{equation}\label{eq:jointHA}
f(\by|\bX,\beta,k)=\prod_{i=1}^n
{\ds \exp\left(\beta \sum_{\ell\,\sim_k i} \delta_{y_i}(y_\ell)\bigg/ k\right)}\Bigg/{\ds 
\sum_{g=1}^G \exp\left(\beta \sum_{\ell\,\sim_k i} \delta_{g}(y_\ell)\bigg/ k\right)}\,.
\end{equation}
Unfortunately, there are drawbacks to this approach, in that, first, the function \eqref{eq:jointHA} 
is not properly normalised (a fact overlooked by \citealp{Holmes:Adams:2002}), and the necessary normalising constant is intractable. Second, the 
full conditional distributions corresponding to this joint distribution are not given by 
\eqref{eq:knn-intuit}. The first drawback is a common occurrence with Boltzmann models and we will
deal with this difficulty in detail in Section \ref{sec:Binf}. At this stage, let us point out
that the most standard approach to this problem is
to use pseudo-likelihood following \cite{besag:york:mollie:1991}, 
as in \cite{heikkinen:hogmander:1994} and \cite{hoeting:etal:1999}, 
but we will show in Section \ref{sec:psudo} that this approximation can give poor results. (See, e.g.,
\cite{friel:pettitt:reeves:wit:2005} for a discussion of this point.) The second and more 
specific drawback implies that \eqref{eq:jointHA} cannot be treated as a pseudo-likelihood 
\citep{Besag:1974,besag:york:mollie:1991}
since, as stated above, the conditional distribution (\ref{eq:knn-intuit})
cannot be associated with any joint distribution. That \eqref{eq:jointHA} misses a normalising
constant can be seen from the special case in which $n=2$, $\by=(y_1,y_2)$ and $G=2$, since
\begin{align*}
\sum_{y_1=1}^2\,\sum_{y_2=1}^2 & \prod_{i=1}^2 {\ds \exp\left(\beta \sum_{\ell\,\sim_k i} 
\delta_{y_i}(y_\ell)\bigg/ k\right)}\Bigg/{\ds \sum_{g=1}^2 \exp\left(\beta 
\sum_{\ell\,\sim_k i} \delta_{g}(y_\ell)\bigg/ k\right)} \\
&= \sum_{y_1=1}^2\,\sum_{y_2=1}^2 {\ds \exp\left(\beta \left[ \delta_{y_1}(y_2) 
+ \delta_{y_2}(y_1) \right] / k \right)}\Bigg/{\ds \left( 1+e^{\beta/k} \right)^2} \\
&= {2\left( 1+e^{2\beta/k} \right)} \bigg/ {\left( 1+e^{\beta/k} \right)^2}\,,
\end{align*}
which is clearly different from $1$ and, more importantly, depends on both $\beta$ and $k$.
We note that the debate about whether or not one should use a proper joint distribution is reminiscent of the opposition
between Gaussian conditional autoregressions (CAR) and Gaussian intrinsic autoregressions in 
\cite{Besag:Kooperberg:1995}, the latter not being associated with any joint distribution.

\subsection{A symmetrised Boltzmann modelling}\label{sec:bobo}
Given these difficulties, we therefore adopt a different strategy 
and define a joint model on the training set as
\begin{equation}
f(\by|\bX,\beta,k)= {\ds \exp\left(\beta \sum_{i=1}^n \sum_{\ell\,\sim_k i} 
\delta_{y_i}(y_\ell)\bigg/ k\right)}\Bigg/{\ds Z(\beta,k)}\,,
\label{eq:knn}
\end{equation}
where $Z(\beta,k)$ is the normalising constant of the distribution. The motivation for this
modelling is that the full conditional distributions corresponding to \eqref{eq:knn}
can be obtained as
\begin{equation}
f(y_i|\by_{-i},\bX,\beta,k)\propto
\exp\left\{ \beta/k \left(\sum_{\ell\,\sim_k i} \delta_{y_i}(y_\ell)
+\sum_{i\sim_k \ell} \delta_{y_\ell}(y_i)\right)\right\}\,,
\label{eq:knn-cond}
\end{equation}
where ${i\sim_k \ell}$ means that the summation is taken over the observations $x_\ell$
for which $x_i$ is a $k$-nearest neighbour. Obviously, these conditional distributions differ
from \eqref{eq:knn-intuit} if only because of the impossibility result mentioned above. 
The additional term in the potential function corresponds to the observations that are not among
the nearest neighbours of $x_i$ but for which $x_i$ is a nearest neighbour. In this model, 
compared with single neighbours, mutual neighbours are given a double weight. This feature is
of importance in that this coherent model defines a new classification criterion that can be treated as a competitor of
the standard \knn~objective function.  Note also that 
the original full conditional \eqref{eq:knn-intuit} is recovered as \eqref{eq:knn-cond} when the 
system of neighbours is perfectly symmetric (up to a factor 2). Once again, the normalising constant
$Z(\beta,k)$ is intractable, except for the most trivial cases.

In the case of unbalanced sampling, that is,
if the marginal probabilities $p_1=\mathbb{P}(y=1),\ldots,p_G=\mathbb{P}(y=G)$
are known and are different from the sampling probabilities $\tilde p_1=n_1/n,\ldots,\tilde p_G=n_G/n$, 
where $n_g$ is the number of training observations arising from class $g$, a natural modification
of this \knn~model is to reweight the neighbourhood sizes by $a_g=p_gn/n_g$. This leads to the
modified model
$$
f(\by|\bX,\beta,k)=
{\ds \exp\left(\beta \sum_i a_{y_i} \sum_{\ell\,\sim_k i} 
\delta_{y_i}(y_\ell)\bigg/ k\right)}\Bigg/{\ds Z(\beta,k)}\,.
$$
This modification is useful in practice when we are dealing with stratified surveys.
In the following, however, we assume that $a_g=1$ for all $g=1,\ldots,G$.

\subsection{Predictive perspective}
When based on the conditional expression \eqref{eq:knn-cond}, the predictive distribution of a new unclassified 
observation $y_{n+1}$ given its covariate $x_{n+1}$ and the training sample $(\by,\bX)$ is, for
$g=1,\ldots,G,$
\begin{equation}
\mathbb{P}(y_{n+1}=g|x_{n+1},\by,\bX,\beta,k)\propto
\exp\left\{\beta/k \left(\sum_{\ell\,\sim_k (n+1)} \delta_{g}(y_\ell)+\sum_{(n+1)\sim_k \ell}
\delta_{y_\ell}(g)\right)\right\}\,,
\label{eq:knn-pred}
\end{equation}
where 
$$
\sum_{\ell\,\sim_k (n+1)} \delta_{g}(y_\ell)\quad\text{ and }\quad \sum_{(n+1)\sim_k \ell} \delta_{y_\ell}(g)
$$
are the numbers of observations in the training dataset from class $g$ among the $k$ nearest neighbours of $x_{n+1}$
and among the observations for which $x_{n+1}$ is a $k$-nearest neighbour, respectively. This predictive distribution
can then be incorporated in the Bayesian inference process by considering the joint posterior of
$(\beta,k,y_{n+1})$ and by deriving the corresponding marginal posterior distribution of $y_{n+1}$.

While this model provides a sound statistical basis for the $k$-nearest-neighour methodology as well as 
a means of assessing the uncertainty of the allocations to classes of unclassified observations, 
and while it corresponds to a true, albeit unavailable, joint distribution, it can be criticised
from a Bayesian point of view in that it suffers from a lack of statistical coherence
(in the sense that the information contained in the sample is not used in the most efficient way)
when multiple classifications are considered. Indeed, the \knn~methodology is invariably 
used in a repeated manner, either jointly on a sample $(x_{n+1},\ldots,x_{n+m})$ or sequentially. 
Rather than assuming simultaneously dependence in the training
sample and independence in the unclassified sample, it would be more sensible to consider the whole
collection of points as issuing from a single joint model of the form given by \eqref{eq:knn},
but with some having their class missing at random. Always reasoning from a Bayesian point of view, 
addressing jointly the inference on the parameters $(\beta,k)$ and on the missing classes 
$(y_{n+1},\ldots,y_{n+m})$---i.e.~assuming {\em exchangeability} between the training and the 
unclassified datapoints---certainly 
makes sense from a foundational perspective as a correct probabilistic evaluation 
and it does provide a better assessment of the uncertainty about the classifications as 
well as about the parameters.

Unfortunately, this more global and arguably more coherent perspective is mostly unachievable if
only for computational reasons, since the set of the missing class vector 
$(y_{n+1},\ldots,y_{n+m})$ is of size $G^m$. It is practically impossible to derive
an efficient simulation algorithm that would correctly approximate the joint probability
distribution of both parameters and classes, especially when the number $m$ of unclassified 
points is large. We will thus adopt the more {\em ad hoc} approach of dealing 
separately with each unclassified point in the analysis, because this simply is the
only realistic way. This perspective can also be justified by the fact that, in realistic machine 
learning set-ups, the unclassified data $(y_{n+1},\ldots,y_{n+m})$ mostly occur in a sequential 
environment with, furthermore, the true value of $y_{n+1}$ being revealed before $y_{n+2}$ is observed.

In the following sections, we mainly consider the case $G=2$ as in \cite{Holmes:Adams:2003},
because this is the only case where we can conduct a full comparison between different
approximation schemes, but we indicate at the end of Section \ref{sec:perfetto} how a Gibbs
sampling approximation allows for a realistic extension to larger values of $G$, as illustrated
in Section \ref{sec:illustre}.

\section{Bayesian inference and the normalisation problem}\label{sec:Binf}

Given the joint model \eqref{eq:knn} for $(y_{1},\ldots,y_{n+1})$, Bayesian inference
can be conducted in a standard manner \citep{Robert:2001}, provided computational
difficulties related to the unavailability of the normalising constant can be solved. 
Indeed, as stressed in the previous section,
from a Bayesian perspective, the classification of unclassified points can be based on the 
marginal predictive (or posterior) distribution of $y_{n+1}$ obtained by integration over the
conditional posterior distribution of the parameters, namely, for $g=1,2,$
\begin{equation}\label{eq:trupred}
\mathbb{P}(y_{n+1}=g|x_{n+1},\by,\bX)=\sum_k\int \mathbb{P}(y_{n+1}=g|x_{n+1},\by,\bX,\beta,k)
\pi(\beta,k|\by,\bX)\,\text{d}\beta\,,
\end{equation}
where $\pi(\beta,k|\by,\bX)\propto f(\by|\bX,\beta,k)\pi(\beta,k)$ is the posterior
distribution of $(\beta,k)$ given the training dataset $(\by,\bX)$. While other choices
of prior distributions are available, we choose for $(k,\beta)$ a uniform prior
on the compact support $\{1,\ldots,K\}\times[0,\beta_{\max{}}]$.
The limitation on $k$ is imposed by the structure of the training dataset in that
$K$ is at most equal to the minimal class size, $\min(n_1,n_2)$, while the 
limitation on $\beta$, $\beta<\beta_{\max{}}$, is customary in Boltzmann models, because of
phase-transition phenomena \citep{Moeller:2003}: when $\beta$ is above a certain value, the 
model becomes "all black or all white", i.e.~all $y_i$'s are either equal to $1$ or to $2$. (This is
illustrated in Figure \ref{fig:constappro1} below by the convergence of the expectation of the
number of identical neighbours to $k$.)
The determination of $\beta_{\max{}}$ is obviously problem-specific and needs to be 
solved afresh for each new dataset since it depends on the topology of the neighbourhood.
It is however straighforward to implement in that a Gibbs simulation of \eqref{eq:knn} for
different values of $\beta$ quickly exhibits the ``black-or-white" features.

\subsection{MCMC steps}\label{sec:mcmstep}
Were the posterior distribution $\pi(\beta,k|\by,\bX)$ available (up to a normalising constant), 
we could design an MCMC algorithm that would produce a Markov chain approximating a sample
from this posterior \citep{Robert:Casella:2004}, for example through a Gibbs sampling scheme 
based on the full conditional distributions of both $k$ and $\beta$. 
However, because of the associated representation \eqref{eq:knn-cond}, the 
conditional distribution of $\beta$ is non-standard and we need to resort to a hybrid sampling
scheme in which the exact simulation from $ \pi(\beta|k,\by,\bX)$ is replaced with a single
Metropolis--Hastings step. Furthermore, use of the full conditional distribution for $k$ can 
impose fairly severe computational constraints. Indeed, for a given value $\beta^{(t)}$, 
computing the posterior $f(\by|\bX,\beta^{(t)},i)\pi(\beta^{(t)},i),$ for $i=1,\ldots,K$, 
requires computations of order
$\text{O}(KnG)$, once again because of the likelihood representation. A faster alternative is 
to use a hybrid step for both $\beta$ and $k$: in this way, we only need to compute the full 
conditional distribution of $k$ for one new value of $k$, modulo the normalising constant. 

An alternative to Gibbs sampling is to use a random walk 
Metropolis--Hastings algorithm: both $\beta$ and $k$ are then updated using random walk proposals. 
Since $\beta\in(0,\beta_{\max})$ is constrained, we first introduce a logistic reparameterisation of $\beta$, 
$$
\beta =\beta_{\max{}}\exp(\theta)\big/ (\exp(\theta)+1)\,,
$$
and then propose a normal random walk on the $\theta$'s, $\theta'\sim\mathcal{N}(\theta^{(t)},\tau^2)$.
For $k$, we use instead a uniform proposal on the $2r$ neighbours of $k^{(t)}$, 
namely $\{ k^{(t)}-r,\ldots,k^{(t)}-1,k^{(t)}+1, \ldots k^{(t)}+r \}\bigcap\{1,\ldots,K\}$.
This proposal distribution with probabiltity density 
$Q_r(k,\cdot)$, with $k'\sim Q_r(k^{(t-1)},\cdot)$,
thus depends on a parameter 
$r\in\{1,\ldots,K\}$ that needs to be calibrated so as to aim at optimal acceptance rates, 
as does $\tau^2$. The acceptance probability in the Metropolis--Hastings algorithm is thus
\begin{eqnarray*}
\rho & = & \frac{f(\by|\bX,\beta',k')\pi(\beta',k')\bigg/ Q_r(k^{(t-1)},k')}
{f(\by|\bX,\beta^{(t-1)},k^{(t-1)})\pi(\beta^{(t-1)},k^{(t-1)})\bigg/ Q_r(k',k^{(t-1)})} \\
     &   & \times~\frac{\exp(\theta')\big/(1+\exp(\theta'))^2}{\exp(\theta^{(t-1)})\big/(1+\exp(\theta^{(t-1)}))^2}\,,
\end{eqnarray*}
where the second ratio is the ratio of the Jacobians due to the reparameterisation. 

Once the Metropolis--Hastings algorithm has produced a satisfactory sequence of $(\beta,k)$'s,
the Bayesian prediction for an unobserved class $y_{n+1}$ associated with $x_{n+1}$ is derived 
from \eqref{eq:trupred}. In fact, if we use a $0-1$ loss function \citep{Robert:2001}
for predicting $y_{n+1}$, namely
$$
\text{L}(\hat y_{n+1},y_{n+1}) = \mathbb{I}_{\hat y_{n+1}\ne y_{n+1}}\,,
$$
the Bayes estimator $\hat y_{n+1}^\pi$ is the most probable class $g$
according to the posterior predictive \eqref{eq:trupred}. The associated measure of
uncertainty is then the posterior expected loss, $\mathbb{P}(y_{n+1}=g|x_{n+1},\by,\bX)$.

Explicit calculation of (\ref{eq:trupred}) is obviously impossible and 
this distribution must be approximated from the MCMC chain $\{(\beta,k)^{(1)},\ldots,(\beta,k)^{(M)}\}$
simulated above, namely by
\begin{equation}
\label{eq:MCpred}
M^{-1}\sum_{i=1}^M {\mathbb{P}}\left(y_{n+1}=g|x_{n+1},\by,\bX,(\beta,k)^{(i)}\right) \,.
\end{equation}

Unfortunately, since \eqref{eq:knn} involves the intractable constant $Z(\beta,k)$, the above schemes
cannot be implemented as such and we need to replace $f$ with a more manageable target. We 
proceed below through three different approaches that try to overcome this difficulty, postponing the comparison
till Section \ref{sec:psudo}.

\subsection{Pseudo-likelihood approximation} 
A first solution, dating back to \cite{Besag:1974}, is to replace the true joint distribution 
with the pseudo-likelihood, defined as
\begin{equation}\label{eq:psudoL}
\hat f(\by|\bX,\beta,k) = \prod_{i=1}^n 
\frac{\ds \exp\left\{ \beta/k \left(\sum_{\ell\,\sim_k i} \delta_{y_i}(y_\ell)
	+\sum_{i\sim_k \ell} \delta_{y_\ell}(y_i)\right) \right\}}
{\ds \sum_{g=1}^2 \exp\left\{ \beta/k \left(\sum_{\ell\,\sim_k i} \delta_{g}(y_\ell)
	+\sum_{i\sim_k \ell} \delta_{y_\ell}(g)\right) \right\}}
\end{equation}
and made up of the product of the (true) conditional distributions associated with \eqref{eq:knn}.
The true posterior distribution $\pi(\beta,k|\by,\bX)$ is then replaced with
$$
\hat\pi(\beta,k|\by,\bX)\propto \hat f(\by|\bX,\beta,k)\pi(\beta,k)\,,
$$
and used as such in all steps of the MCMC algorithm drafted above.
The predictive distribution $\mathbb{P}(y_{n+1}=g|x_{n+1},\by,\bX)$ 
is correspondingly approximated by \eqref{eq:MCpred}, based on the pseudo-sample thus produced.

While this replacement of the true distribution with the pseudo-likelihood approximation
induces a bias in the estimation of $(k,\beta)$ and 
in the predictive performance of the Bayes procedure, it has been intensively used in the past, if
only because of its availability and simplicity. For instance, \cite{Holmes:Adams:2003} built their
pseudo-joint distribution on such a product (with the difficulty that the components of the product
were not true conditionals). As noted in Friel and Pettitt
(\citeyear{Friel:Pettitt:2004}), pseudo-likelihood estimation can
be very misleading and we will describe its performance in more detail in Section \ref{sec:psudo}.
(To the best of our knowledge, this Bayesian evaluation has not been conducted before.)

\begin{figure}
\centerline{\includegraphics[width=220pt]{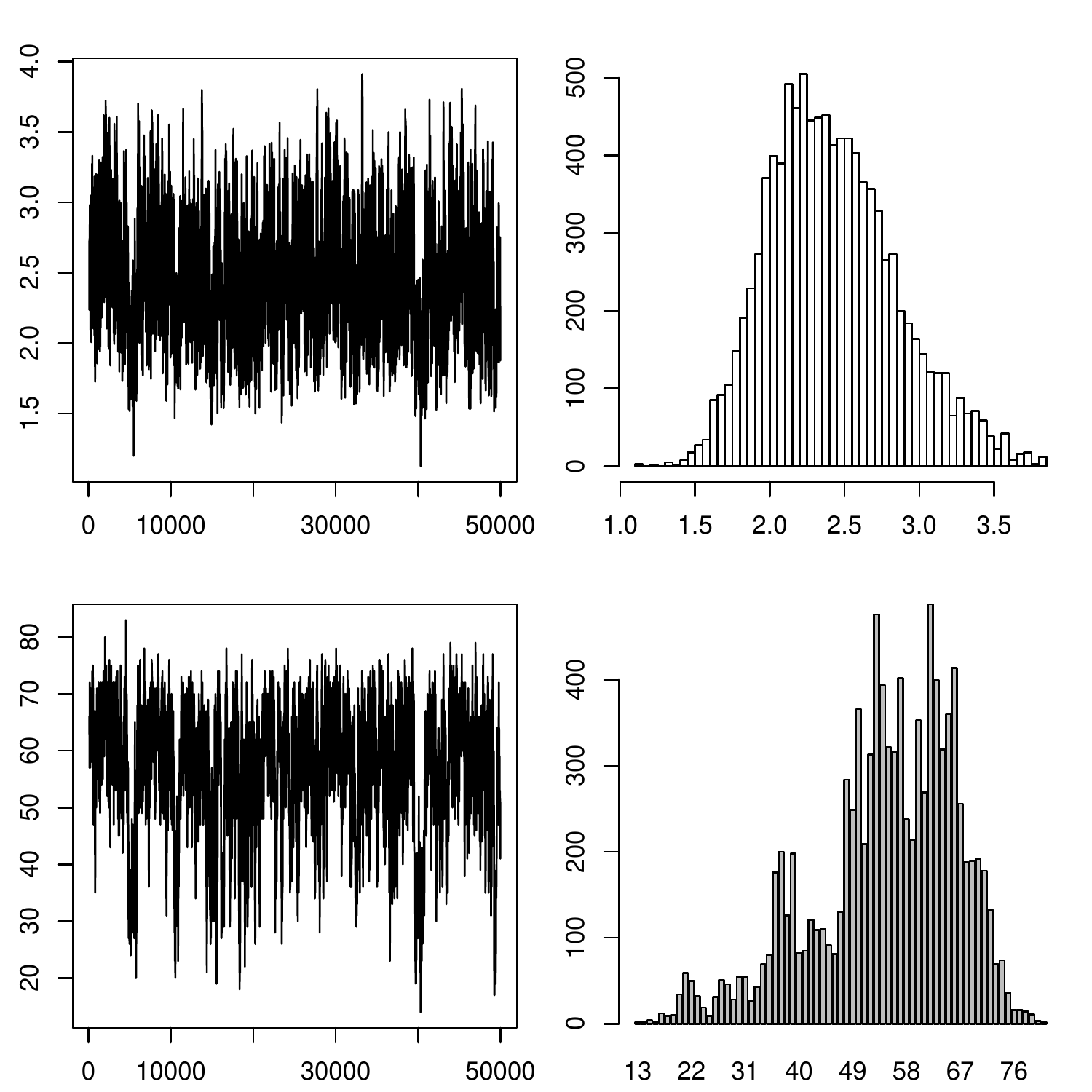}}
\caption{
Output of a random walk Metropolis--Hastings algorithm based on the pseudo-likelihood 
approximation of the normalising constant for $50,000$ iterations, with a $40,000$ iteration
burn-in stage, and $\tau^2=0.05$, $r=3$. {\em (top)} sequence and marginal histogram for $\beta$ 
when $\beta_{\max{}}=4$ and {\em (bottom)} sequence and marginal barplot for $k$.
\label{pseudoo}}
\end{figure}

As illustrated on Figure \ref{pseudoo} for Ripley's benchmark data, the 
random walk Metro\-polis--Hastings algorithm 
detailed above performs satisfactorily with the pseudo-likelihood approximation, 
even though the mixing is slow (cycles can be spotted on the
bottom left graph). On that dataset,
the pseudo-maximum--i.e., the maximum of \eqref{eq:psudoL}--is 
achieved for $\hat k=53$ and $\hat \beta=2.28$. If we use the last $10,000$ iterations of this MCMC run,
the prediction performance of \eqref{eq:MCpred} is such that the error rate on the test set of $1000$
points is $8.7\%$. Figure \ref{pseudoo} also indicates how limited the information is about $k$.
(Note that we settled on the value $\beta_{\max{}}=4$ by trial-and-error.)

\subsection{Path sampling}\label{sec:path}

A now-standard approach to the estimation of normalising constants is {\em path sampling}, 
described in \cite{Gelman:Meng:1998} (see also \citealp{Chen:Shao:Ibrahim:2000}), 
and derived from the \cite{Ogata:1989} method, in which the ratio of two normalising constants, 
$Z(\beta^\prime,k)/Z(\beta,k)$, can be decomposed as an integral to be approximated by
Monte Carlo techniques. 

The basic derivation of the path sampling approximation is that, if 
$$
S(\by) = \sum_i\sum_{\ell\,\sim_k i} \delta_{y_i}(y_\ell)/k\,,
$$
then
$$
{Z}(\beta,k)=\sum_{\by}\exp\left[\beta S(\by) \right]\\
$$
and
\begin{eqnarray*}
\frac{\partial{Z}(\beta,k)}{\partial \beta}
& = & \sum_{\by} S(\by) \exp[\beta S(\by) ]\\
& = & {Z}(\beta,k)\,\sum_{\by} S(\by)\,{\exp(\beta S(\by))}\big/{{Z}(\beta,k)} \\
& = & {Z}(\beta,k)\,\mathbb{E}_{\beta}[S(\by)]\,.
\end{eqnarray*}
Therefore, the ratio $Z(\beta,k)/{Z}(\beta^\prime,k)$ can be derived from an integral, 
since
$$
\log\left\{ {Z}(\beta,k)/{Z}(\beta^\prime,k)\right\}
=\int_{\beta}^{\beta^\prime} \mathbb{E}_{u,k} [S(\by)]\,\text{d}u\,,
$$
which is easily evaluated by a numerical approximation.

The practical drawback with this approach is that each time a new ratio is to be 
computed, that is, at each step of a hybrid Gibbs scheme or of a Metropolis--Hastings proposal, 
an approximation of the 
above integral needs to be produced. A further step is thus necessary for path sampling to 
be used: we approximate the function ${Z}(\beta,k)$ only once for each value of $k$ and for a 
few selected values of $\beta$, and later we use numerical interpolation to extend the
function to other values of $\beta$. Since the function ${Z}(\beta,k)$ is very smooth, the
degree of additional approximation is quite limited. Given that this approximation is only
to be computed once, the resulting Metropolis-Hastings algorithm is very fast, as well as being
efficient if enough care is taken with the approximation by checking that the slope of
${Z}(\beta,k)$ is sufficiently smooth from one value of $\beta$ to the next.
(We stress however that the computational cost required to produce those approximations is fairly high,
because of the joint approximation in $(\beta,k)$.)

\begin{figure}
\centerline{\includegraphics[width=100pt]{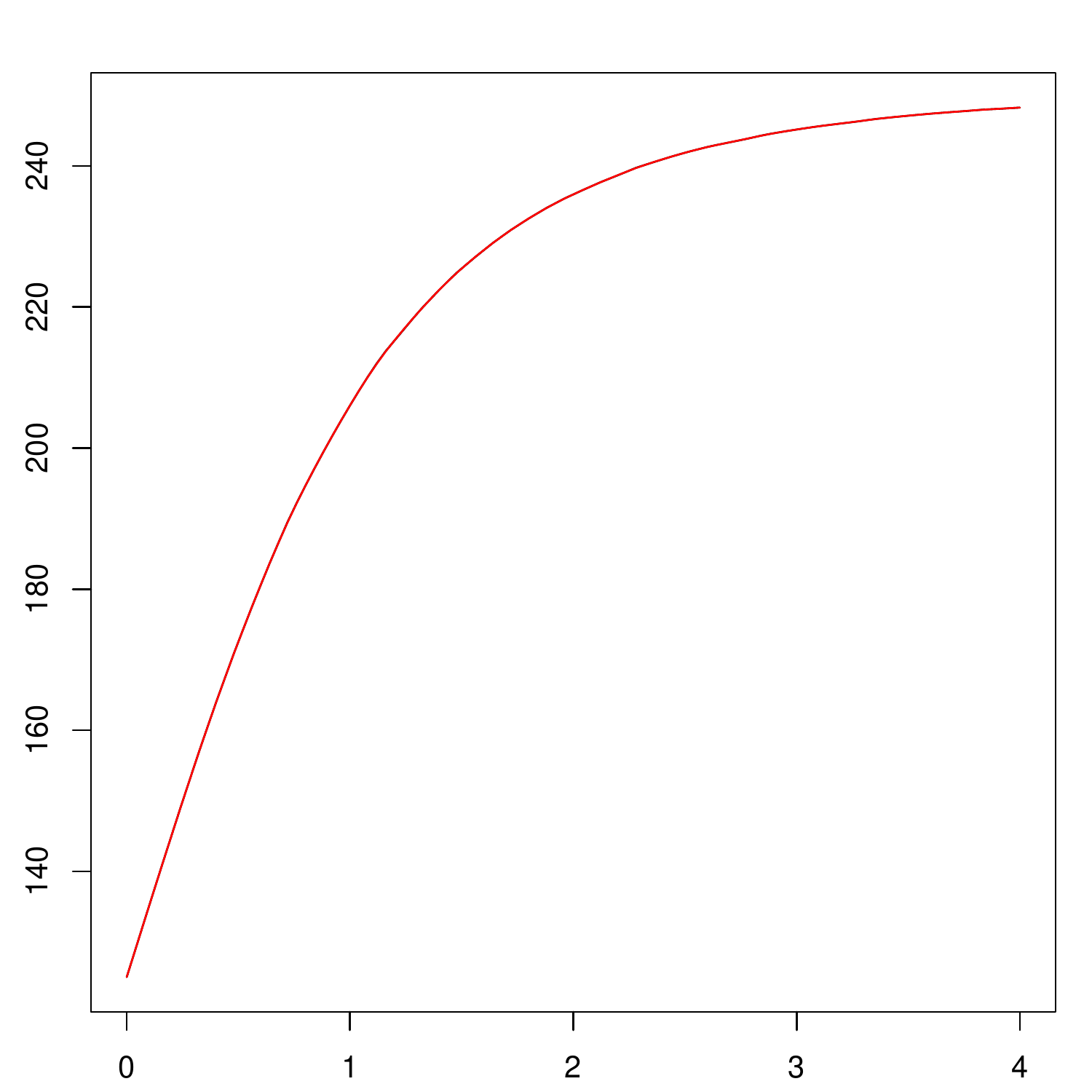}\hglue1truecm\includegraphics[width=100pt]{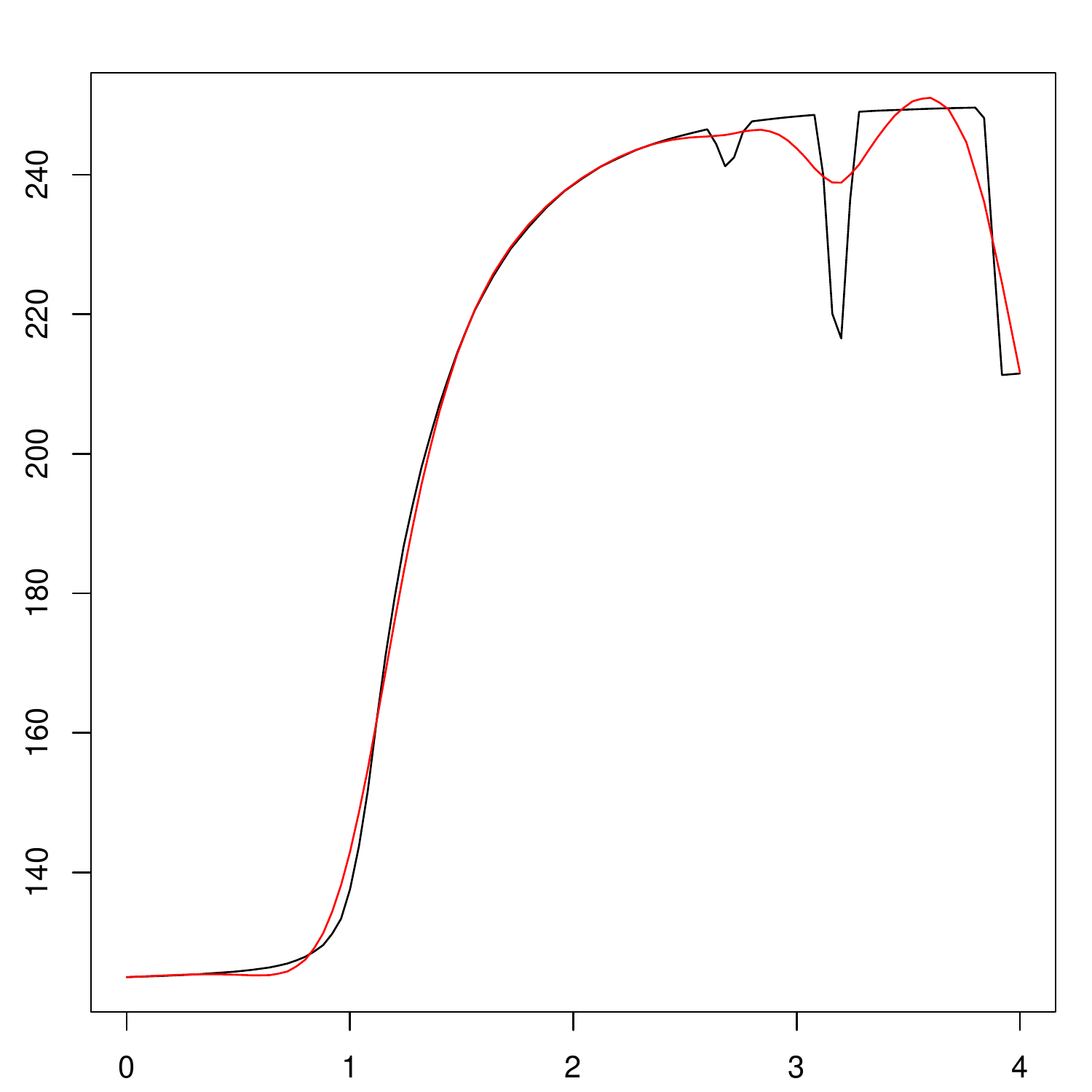}}

\caption{
Approximation of the expectation $\mathbb{E}_{\beta,k} [S(\by)]$ for Ripley's benchmark,
where the $\beta$'s are equally spaced between $0$ and $\beta_{\max{}}=4$, and 
for $k=1$ {\em (left)} and $k=125$ {\em (right)}
($10^4$ iterations with $500$ burn-in steps for each value of $(\beta,k)$). On these graphs, the
black curve is based on linear interpolation of the expectation and the red curve on
second-order spline interpolation. 
\label{fig:constappro1}}
\end{figure}

We illustrate this approximation using Ripley's benchmark dataset.
Figure \ref{fig:constappro1} provides the approximated expectations $\mathbb{E}_{\beta,k} [S(\by)]$ for 
a range of values of $\beta$ and for two values of $k$. Within the
expectation, the $\by$'s are simulated using a systematic scan Gibbs
sampler, because using the perfect sampling scheme elaborated below in Section \ref{sec:perfetto}
makes little sense when only one expectation needs to be computed. As seen from this
comparative graph, when $\beta$ is small, the Gibbs sampler gives good mixing performance,
while, for larger values, it has difficulty in converging, as illustrated by the poor fit on
the right-hand plot when $k=125$. The explanation is that the model is getting closer to the
phase-transition boundary in that case.

\begin{figure}
\centerline{\includegraphics[width=160pt]{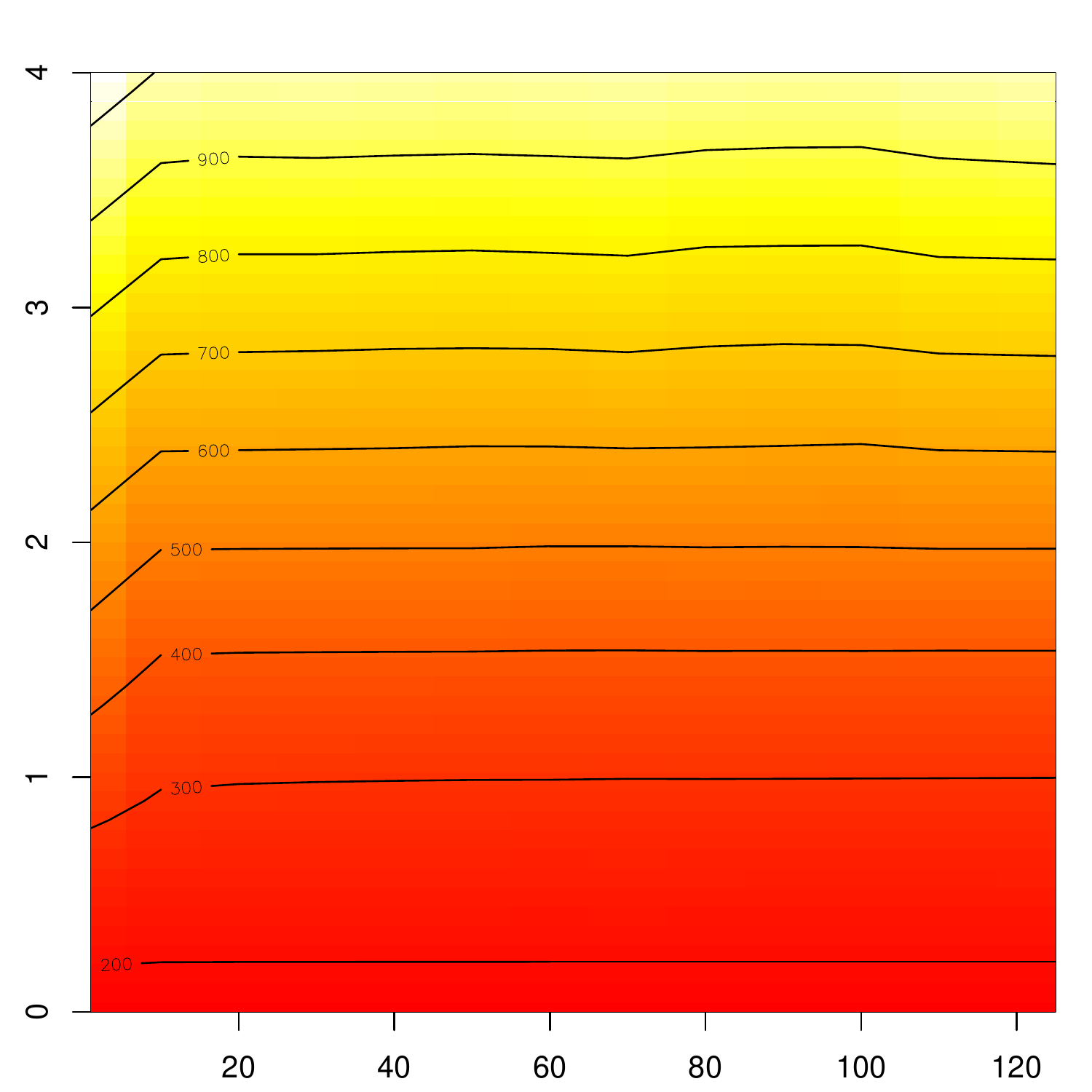} }

\caption{
Approximation of the normalising constant $Z(\beta,k)$ for Ripley's dataset where
the $\beta$'s are equally spaced between $0$ and $\beta_{\max{}}=4$,
and $k=1,10,20,\ldots,110,125$ (based on $10^4$ Monte Carlo iterations with $500$ burn-in steps, 
and bilinear interpolation).
\label{fig:constappro2}}
\end{figure}

For the approximation of ${Z}(\beta,k)$, we use the fact that $\mathbb{E}_{\beta,k}[S(\by)]$
is known when $\beta=0$, namely $\mathbb{E}_{0,k}[S(\by)]=n/2$. We can
thus represent $\log\{{Z}(\beta,k)\}$ as 
$$
n\log 2 + \int_0^\beta \mathbb{E}_{u,k} [S(\by)]\,\text{d}u
$$
and use numerical integration to approximate the integral. As shown on Figure \ref{fig:constappro2},
which uses a bilinear interpolation based on a $50\times 12$ grid of values of $(\beta,k)$,
the approximated constant $Z(\beta,k)$ is mainly constant in $k$. 

\begin{figure}
\centerline{\includegraphics[width=220pt]{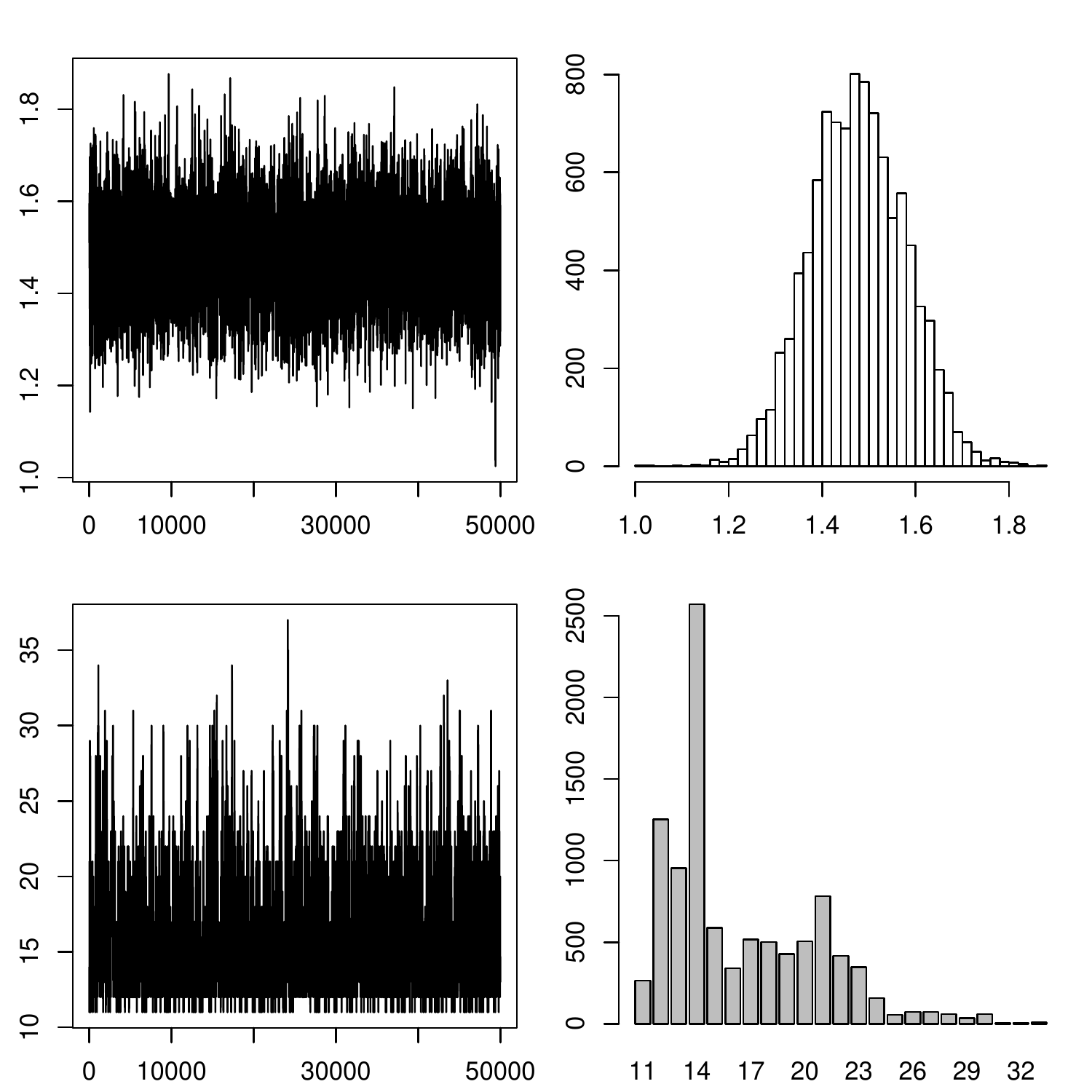}}
\caption{
Output of a random walk Metropolis--Hastings algorithm based on the path sampling approximation of the
normalising constant for $50,000$ iterations, with a $40,000$ iteration burn-in stage and $\tau^2=0.05$,
$r=3$. {\em (top)} sequence and marginal histogram for $\beta$ when $\beta_{\max{}}=4$
and {\em (bottom)} sequence and marginal barplot for $k$.
\label{psycho}}
\end{figure}

Once $Z(\beta,k)$ has been approximated, 
we can use the genuine MCMC algorithm of
Section \ref{sec:mcmstep} fairly easily, the main cost of this approach being thus in the approximation 
of $Z(\beta,k)$. Figure \ref{psycho} illustrates the output of the MCMC sampler for Ripley's benchmark, 
to be compared with Figure \ref{pseudoo}. A first item of interest is that the chain mixes much more
rapidly(in terms of iterations) than its pseudo-likelihood counterpart. A more important point is 
that the range and shape of the
approximations to both marginal posterior distributions differ widely between the two methods, a feature discussed
in Section \ref{sec:psudo}. When this output of the MCMC sampler is used for prediction purposes in \eqref{eq:MCpred},
the error rate for Ripley's test set is equal to $8.5\%$. 

\newcommand{\bz}{\mathbf{z}}
\subsection{Perfect sampling implementation and Gibbs approximation}\label{sec:perfetto}
A completely different approach to handling missing normalising constants
has been developed recently by \cite{Moeller:Pettitt:Reeves:2006} and is based 
on an auxiliary variable idea. If we
introduce an auxiliary variable $\bz$ on the same state space
as $\by$, with arbitrary conditional density $g(\bz|\beta,k,\by)$, and 
if we consider the joint posterior 
$$
\pi(\beta,k,\bz|\by) \propto \pi(\beta,k, \bz, \by) 
= g(\bz|\beta,k,\by)\times f(\by|\beta,k) \times \pi(\beta,k)\,,
$$
then simulating $(\beta,k,\bz)$ from this posterior is equivalent to simulating $(\beta,k)$
from the original posterior since $\bz$ integrates out. If we now run a Metropolis-Hastings
algorithm on this augmented scheme, with $q_1$ an arbitrary proposal density on $(\beta,k)$ and
with
$$
q_2(\beta',k',\bz'|\beta,k,\bz)=q_1(\beta',k'|\beta,k,\by)f(\bz'|\beta',k') \,,
$$
as the joint proposal on $(\beta,k,\bz)$ (i.e., simulating $\bz$ directly from the likelihood),
the Metropolis-Hastings ratio associated with $q_2$ is
\begin{align*}
{\left(\frac{{Z}(\beta,k)}{{Z}(\beta',k)}\right)}
&\left(\frac{\exp\left({\beta'}S(\by)/k'\right)\pi(\beta',k')}{\exp\left(\beta S(\by)/k\right)
\pi(\beta,k)}\right)\left(\frac{g(\bz'|\beta',k',\by)}{g(\bz|\beta,k,\by)}\right)\\
&\times\left(\frac{q_1(\beta,k|\beta',k,\by)\exp\left(\beta S(\bz)/k\right)}
{q_1(\beta',k'|\beta,k,\by)\exp\left({\beta'}S(\bz)/k'\right)}\right)
{\left(\frac{{Z}(\beta',k')}{{Z}(\beta,k)}\right)}\,,
\end{align*}
which means that the constants $Z(\beta,k)$ and $Z(\beta',k')$ cancel out.
The method of \cite{Moeller:Pettitt:Reeves:2006} can thus be calibrated by the choice of
the artificial target $g(\bz|\beta,k,\by)$ on the auxiliary variable $\bz,$ and the authors
advocate the choice
$$
g(\bz|\beta,k,\by)
=\exp\left(\hat{\beta}S(\bz)/\hat k\right)\big/{Z}(\hat\beta,\hat k),
$$
as reasonable, where $(\hat\beta,\hat k)$ is a preliminary estimate,
such as the maximum pseudo-likelihood estimate. While we follow this recommendation,
we stress that the choice of $(\hat\beta,\hat k)$ is paramount for good performance
of the algorithm, as explained below. 
The alternative of setting a target $g(\bz|\beta,k,\by)$ that truly depends on $\beta$ and $k$ is
appealing but faces computational difficulties in that the most natural proposals involve normalising
constants that cannot be computed.

Obviously, this approach of \cite{Moeller:Pettitt:Reeves:2006} 
also has a major drawback, namely that the auxiliary variable $\bz$ must
be simulated from the distribution $f(\bz|\beta,k)$ itself. However, there have been many
developments in the simulation of Ising models, from \cite{Besag:1974} to \cite{Moeller:Waagepetersen:2003}, and 
the particular case $G=2$ allows for exact simulation of $f(\bz|\beta,k)$ using perfect sampling.
We refer the reader to \cite{Haggstrom:2002}, \cite{Moeller:2003}, \cite{Moeller:Waagepetersen:2003} 
and \citeauthor{Robert:Casella:2004} (2004, 
Chapter 13) for details of this simulation technique and for a discussion of its limitations. 
Without entering into technical details, we comment
that, in the case of model \eqref{eq:knn} with $G=2$, there also exists a monotone implementation
of the Gibbs sampler that allows for a practical implementation of the perfect sampler 
\citep{Kendall:Moeller:2000,Berthelsen:Moeller:2003}. More precisely, we can use a 
coupling-from-the-past strategy \citep{Propp:Wilson:1998}: in this setting, starting from the
saturated situations in which the components of $\bz$ are either all equal to $1$ or all equal
to $2$, it is sufficient to monitor both associated chains further and further into the past
until they coalesce by time $0$. The sandwiching property of \cite{Kendall:Moeller:2000}
and the monotonicity of the Gibbs sampler ensure that all other chains associated with arbitrary
starting values for $\bz$ will also have coalesced by then.
The only difficulty with this perfect sampler is
the phase-transition phenomenon, which means that, for very large values of $\beta$, the 
convergence performance of the coupling from the past sampler deteriorates quite rapidly,
a fact also noted in \cite{Moeller:Pettitt:Reeves:2006} for the Ising model. We overcome this
difficulty by using an additional accept-reject step based on smaller values of $\beta$ that 
avoids this explosion in the computational time.

\begin{figure}
\centerline{\includegraphics[width=220pt]{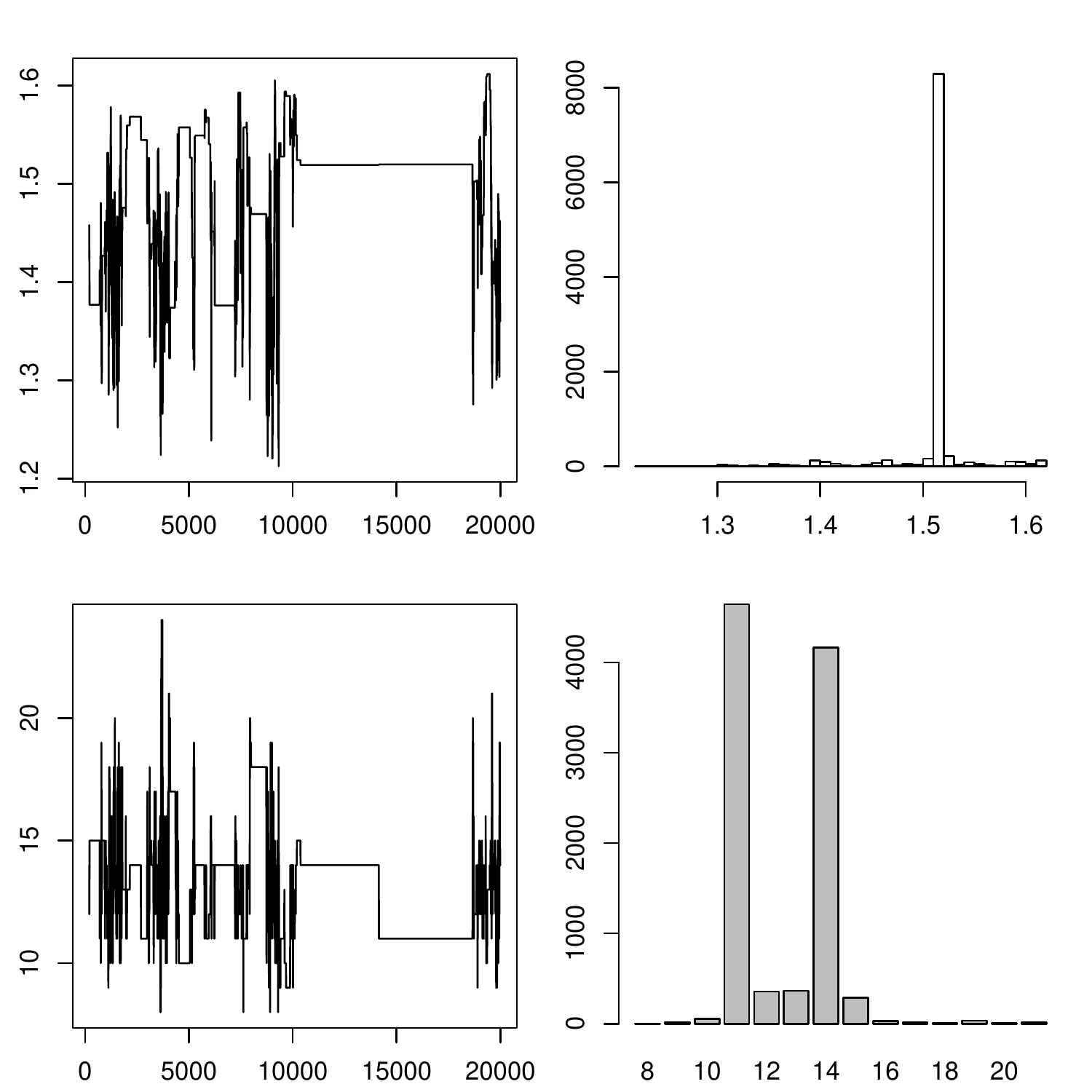}}
\caption{
Output of a random walk Metropolis--Hastings algorithm based on the perfect sampling elimination of the
normalising constant for a pseudo-likelihood plug-in estimate $(\hat k,\hat \beta)=(53,2.28)$ 
and $20,000$ iterations, with a $10,000$ burn-in stage, $\beta_{\max{}}=4$ and $\tau^2=0.05$,
$r=3$: {\em (top)} sequence and marginal histogram for $\beta$ 
and {\em (bottom)} sequence and marginal barplot for $k$.
\label{imperfecto}}
\end{figure}

\begin{figure}
\centerline{\includegraphics[width=220pt]{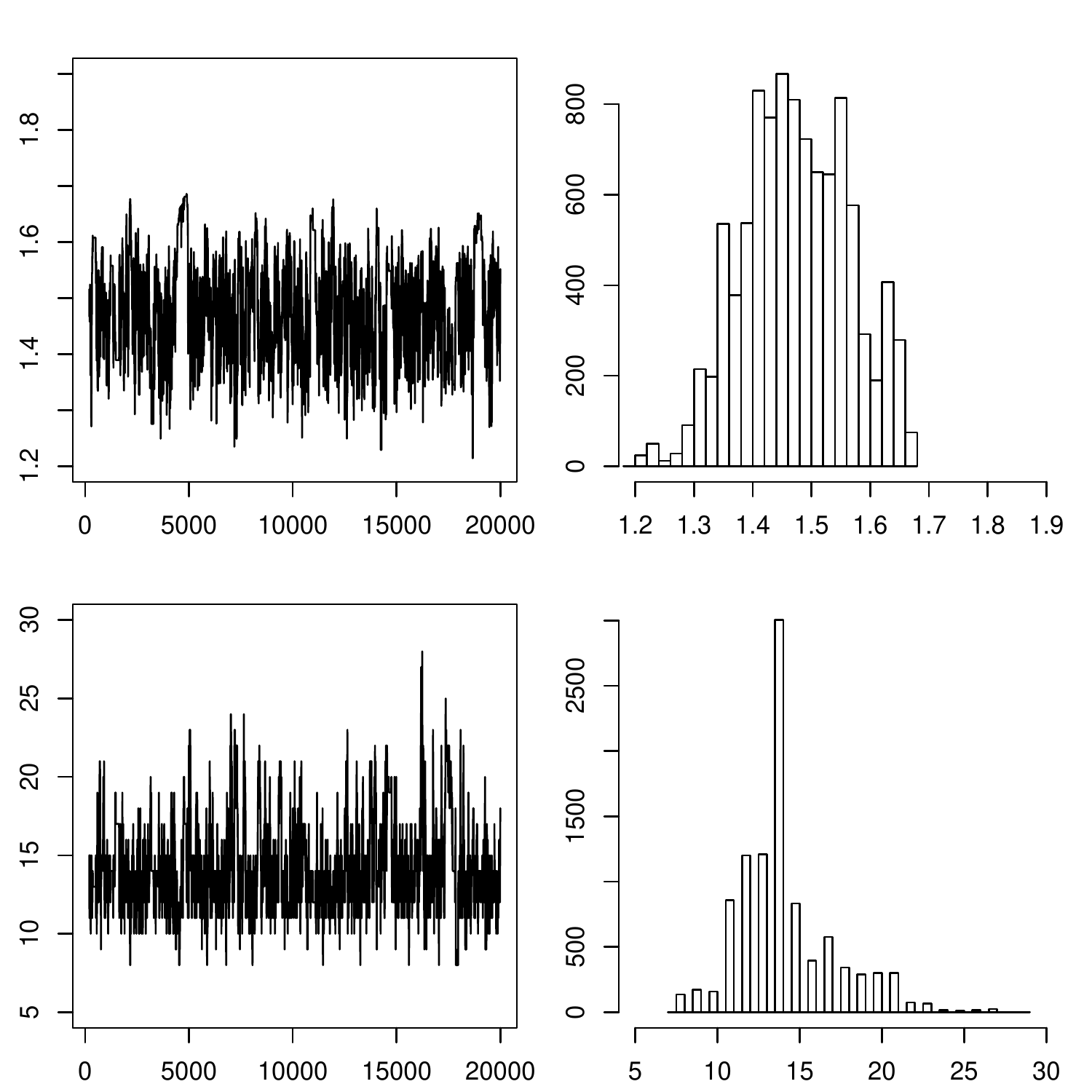}
\includegraphics[width=220pt]{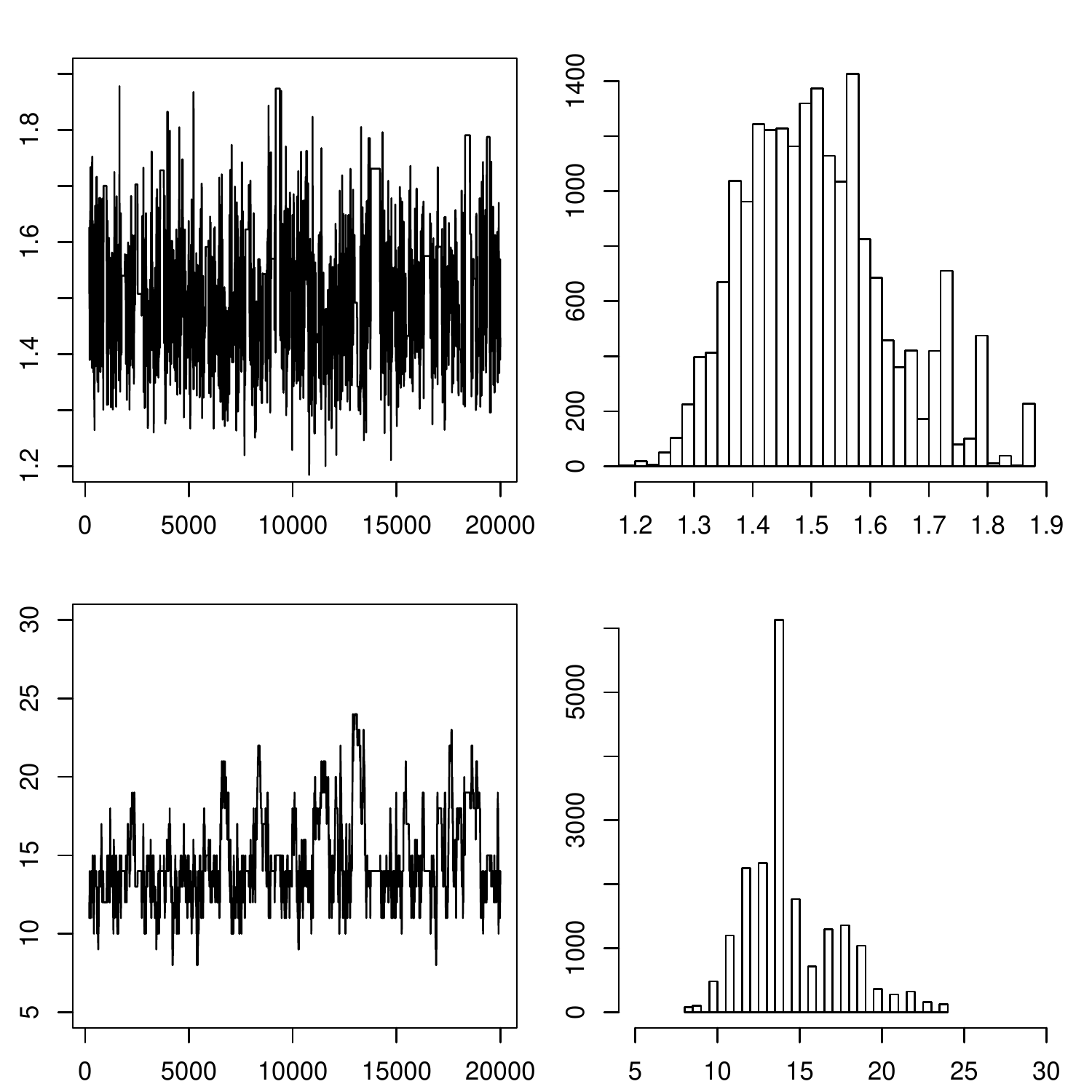}}
\caption{
Comparison of {\em (left)} the output of a random walk Metropolis--Hastings algorithm based on perfect sampling and of 
{\em (right)} the output of its Gibbs approximation
for a plug-in estimate $(\hat k,\hat \beta)=(13,1.45)$ 
and $20,000$ iterations, with a $10,000$ burn-in stage and $\tau^2=0.05$,
$r=3$: {\em (top)} sequence and marginal histogram for $\beta$ 
and {\em (bottom)} sequence and marginal barplot for $k$.
\label{perfecto}}
\end{figure}

As shown on Figure \ref{imperfecto}, a poor choice for $(\hat\beta,\hat k)$ leads to
very unsatisfactory performance with the algorithm. Starting from the pseudo-likelihood
estimate and using this very value for the plug-in value $(\hat\beta,\hat k)$, we obtain
a Markov chain with a very low energy and a very high rejection rate. However, use of the estimate 
$(\hat k,\hat \beta)=(13,1.45)$ resulting 
from this poor run does improve considerably the performance of the algorithm, as shown
by Figure \ref{perfecto}. In this setting, the predictive error rate on the test dataset
is equal to $0.084$.

While this elegant solution based on an auxiliary variable completely removes the issue of the normalising constant, it 
faces several computational difficulties. First, as noted above, the choice of the artificial target $g(\bz|\beta,k,\by)$ 
is driving the algorithm and plug-in estimates need to be reassessed periodicaly. Second, perfect simulation from the 
distribution $f(\bz|\beta,k)$ is extremely costly and may fail if $\beta$ is close to the phase-transition 
boundary. Furthermore, the numerical value of this critical point is not known beforehand. Finally, the extension of the 
perfect sampling scheme to more than $G=2$ classes has not yet been achieved.

For these different reasons, we advocate the substitution of a Gibbs sampler for the above perfect sampler in order to
achieve manageable computing performance. If we replace the perfect sampling step with $500$ (complete) iterations of 
the corresponding generic Gibbs sampler on $\bz$, the computing time is linear in the number $n$ of observations and the results are virtually the same. One has to remember that the simulation of $\bz$ is of second-order with respect to the original problem
of simulating the posterior distribution of $(\beta,k)$, since $\bz$ is an auxiliary variable introduced to overcome the
computation of the normalising constant. Therefore, the additional uncertainty induced by the use of the Gibbs sampler is
far from severe. Figure \ref{perfecto} compares the Gibbs solution with the perfect sampling implementation
and it shows how little loss is incurred by the use of the less expensive Gibbs sampler, while the gain in computing time
is enormous. For $50,000$ iterations, the time required to run the Gibbs sampler is approximately $20$ minutes, compared
with more than a week for the corresponding perfect sampler (under the same {\sf C} environment on the same machine).

\subsection{Evaluation of the pseudo-likelihood approximation}\label{sec:psudo}

Given that the above alternatives can all be implemented for small values of $n$, it is of 
direct interest to compare them in order to evaluate the effect of the pseudo-likelihood
approximation. As demonstrated in the previous section,
using Ripley's benchmark with a training set of $250$ points, we are indeed
able to run a perfect sampler over the range of possible $\beta$'s,
and this implementation gives a sampler in which the only approximation is due to running an MCMC
sampler (a feature common to all three versions).

\begin{figure}
\centerline{\includegraphics[width=250pt]{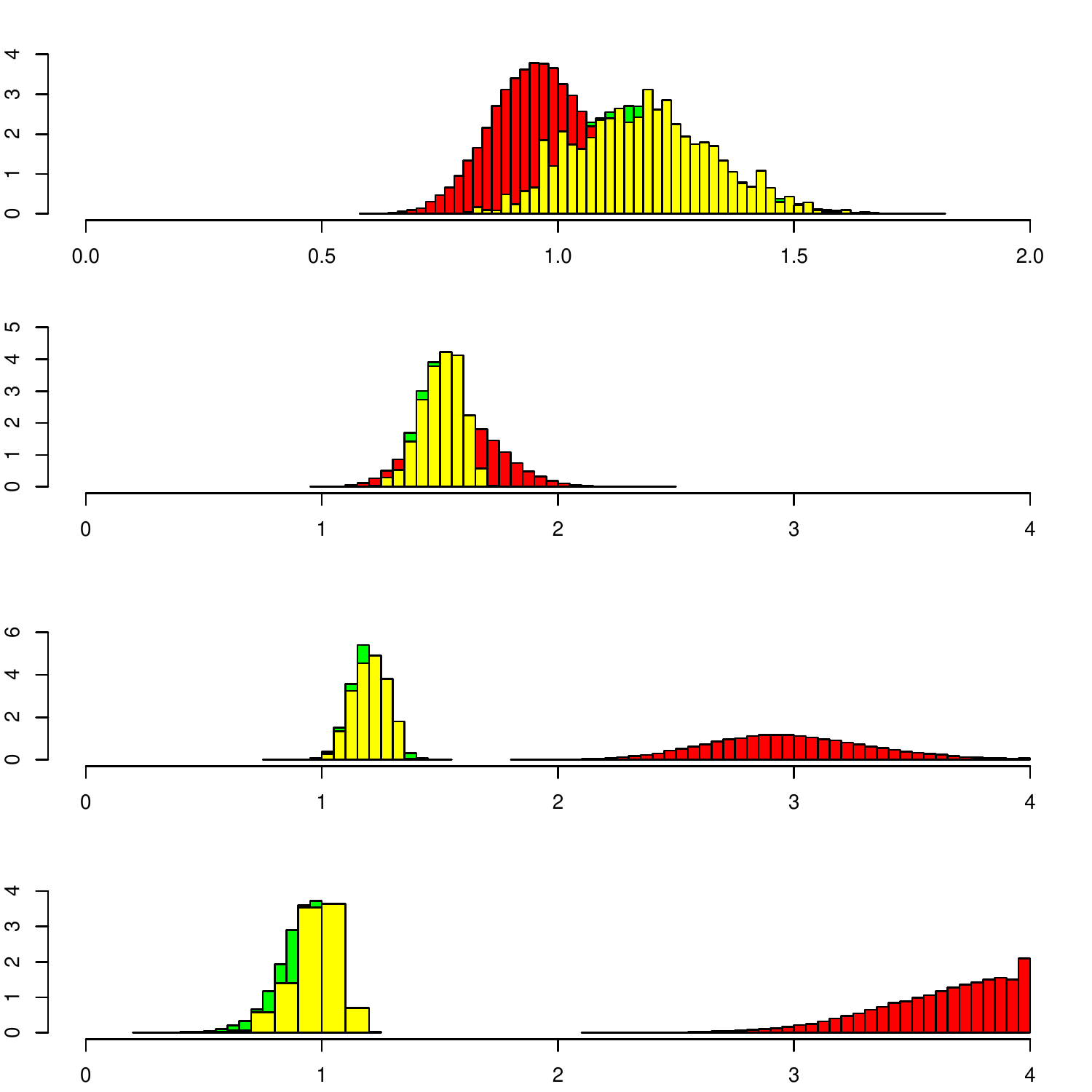}}
\caption{
Comparison of the approximations to the posterior distribution of $\beta$ based on
the pseudo {\em (red)}, the path {\em (green)} and the perfect {\em (yellow)}
schemes for Ripley's benchmark and $k=1,10,70,125$, for $20,000$ iterations and $10,000$ burn-in.
\label{pseudoA}}
\end{figure}

\begin{figure}
\centerline{\includegraphics[width=250pt]{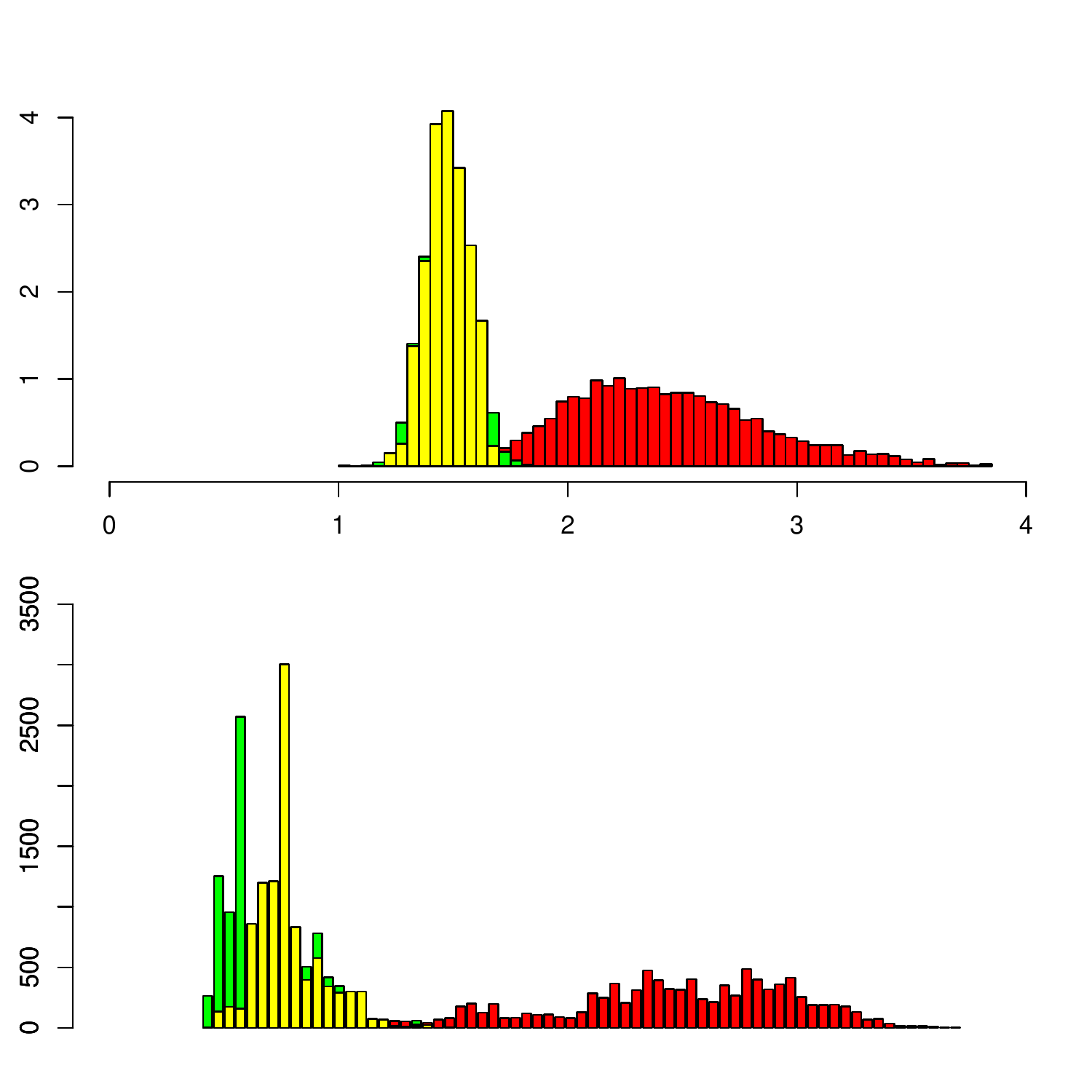}}
\caption{
Comparison of posterior distributions of $\beta$ {\em (top)} and $k$ 
{\em (bottom)} as represented in Figure \ref{pseudoo} for the pseudo-likelihood
approximation, in Figure \ref{psycho} for the path sampling approximation
and in Figure \ref{perfecto} for the perfect sampling approximation.
\label{pseudoB}}
\end{figure}

Histograms, for the same dataset, of simulated $\beta$'s, conditional or unconditional, on $k$
show gross misrepresentation of the samples produced by the
pseudo-likelihood approximation; see Figures \ref{pseudoA} and \ref{pseudoB}. (The comparison for a fixed value of $k$ was obtained directly
by setting $k$ to a fixed value in all three approaches and running the corresponding MCMC
algorithms.)
It could of course be argued that the defect lies with the path sampling evaluation of the constant,
but this approach strongly coincides with the perfect sampling implementation, as showed on both figures.
There is thus a fundamental discrepancy in using the pseudo-likelihood approximation; in other words,
the pseudo-likelihood approximation defines a clearly different posterior distribution on $(\beta,k)$.

As exhibited on Figure \ref{pseudoA}, the larger $k$ is, the worse is this discrepancy, whereas Figure \ref{pseudoB}
shows that both $\beta$ and $k$ are significantly overestimated by the pseudo-likelihood approximation. (It is
quite natural to find such a correlation between $\beta$ and $k$ when we realise that the likelihood
depends mainly on $\beta/k$.) We can also note that the correspondence between path and perfect approximations is not
absolute in the case of $k$, a difference that may be attributed to slower convergence in one or both
samplers.

\begin{figure}
\centerline{\includegraphics[width=250pt]{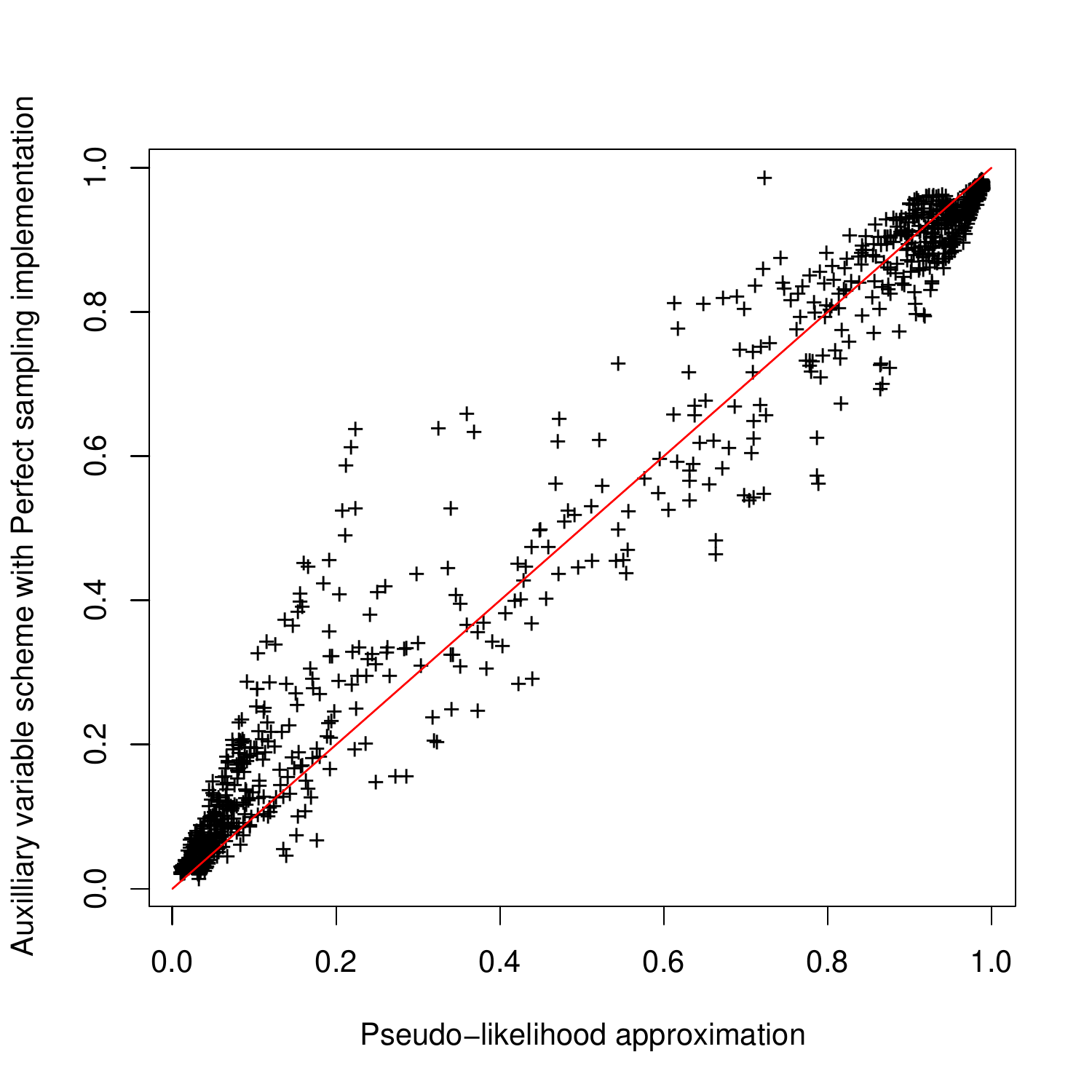}}
\caption{
Comparison of the class probabilities $\mathbb{P}(y=1|x,\by,\bX)$
estimated at each point of the testing sample.
\label{proba-compa}}
\end{figure}

In order to assess the comparative predictive properties of both approaches, we also provide a comparison
of the class probabilities $\mathbb{P}(y=1|x,\by,\bX)$ estimated at each point of the test sample.
As shown by Figure \ref{proba-compa}, the predictions are quite different for values in the middle of the range, with no clear bias
direction in using pseudo-likelihood as an approximation. Note that the discrepancy may be substantial and may result
in a large number of different classifications.

\section{Illustration on real datasets}\label{sec:illustre}

In this Section, we illustrate the behaviour of the proposed methodology on some benchmark datasets.

We first describe the calibration of the algorithm used on each dataset.
As starting value for the Gibbs approximation in the M{\o}ller scheme, we use
the maximum pseudo-likelihood estimate. The Gibbs sampler is iterated 500 times
as an approximation to the perfect sampling step.
After 10,000 iterations, we modify the plug-in estimate using the current average and
then we run 50,000 more iterations of the algorithm. 

The first dataset is borrowed from the {\sf MASS} library of {\sf R}. It consists in the
records of 532 Pima Indian women who were tested by the U.S. National Institute of Diabetes and
Digestive and Kidney Diseases for diabetes. Seven quantitative covariates were recorded, along with
the presence or absence of diabetes. The data are split at random into a training set of
200 women, including $68$ diagnosed with diabetes, and a test set of the remaining 
332 women, including $109$ diagnosed with diabetes. The performance for various values of
$k$ on the test dataset is given in Table \ref{tab:knn-indian}. If we use a standard leave-one-out 
cross-validation for selecting $k$ (using only the training dataset), then there are $10$ consecutive
values of $k$ leading to the same error rate, namely the range $57$--$66$.

\begin{table}[hbtp]
\begin{center}
\begin{tabular}{cc}
 $k$  & Misclassification \\
      & error rate \\
\hline
 1    & 0.316 \\
 3    & 0.229 \\
 15   & 0.226 \\
 31   & 0.211 \\
 57   & 0.205 \\
 66   & 0.208 \\
\hline
\end{tabular}
\end{center}
\caption{Performance of \knn~methods on the Pima Indian test dataset.
\label{tab:knn-indian}}
\end{table}

\begin{figure}
\centerline{\includegraphics[width=220pt]{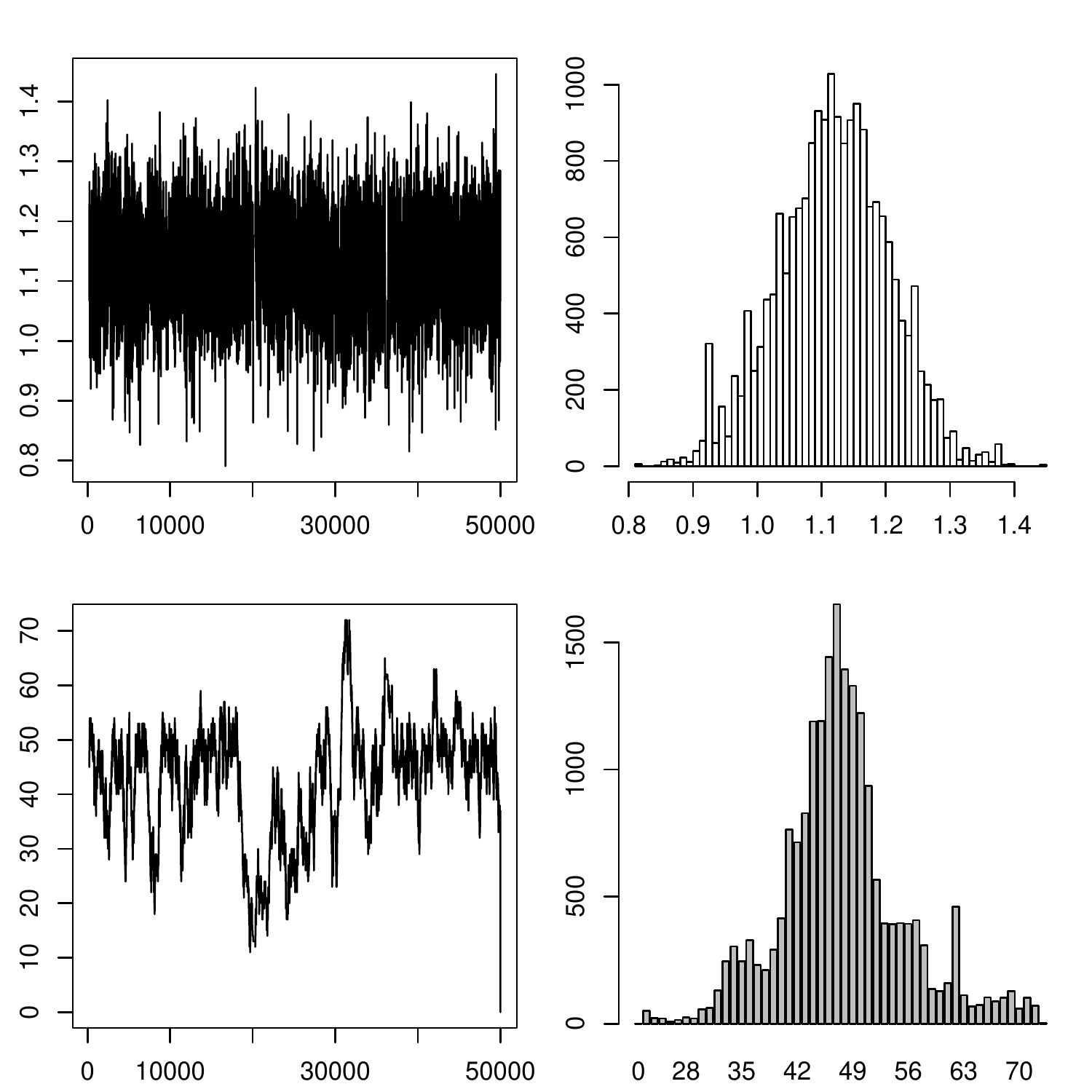}}
\caption{
Pima Indian diabetes study based on $50,000$ iterations of the Gibbs-M{\o}ller sampling scheme with
$\tau^2=0.05$, $r=3$, $\beta_{\max}=4$, and $K=68$.
\label{perfindian}}
\end{figure}

The results are provided in Figure \ref{perfindian}. 
Note that the simulated values of $k$ tend to avoid the region found by the cross-validation
procedure. One possible reason for this discrepancy is that, as noted in Section \ref{sec:bobo},
the likelihood for our joint model is not directly equivalent to the \knn~objective function, since mutual neighbours are
weighted twice as heavily as single neighbours in this likelihood.
Over the final $20,000$ iterations, the prediction error is $0.209$, quite in line with the \knn~solution
in Table \ref{tab:knn-indian}.

To illustrate the ability of our method to consider more than two classes, we also used the benchmark
dataset \textsf{forensic glass fragments}, studied in \cite{Ripley:1994}.
This dataset  involves nine covariates and six classes some of which are rather rare. Following the recommendation
made in \cite{Ripley:1994}, we coalesced some classes to reduce the number of classes to four. We then randomly
partitioned the dataset to obtain 89 individuals in the training dataset and 96 in the testing dataset.
Leave-one-out cross-validation leads us to choose the value $k=17$. The error rate of the 
$17$-nearest-neighbour procedure
on the test dataset is $0.35$, whereas, using our procedure, we obtain an error rate of $0.29$.
The substantial gain from using our approach can be partially explained by the fact that the value of $k$ chosen by the cross-validation procedure
is much larger than those explored by our MCMC sampler. 

\section{Conclusions}
While the probabilistic background to a Bayesian analysis of \knn~methods was initiated by 
\cite{Holmes:Adams:2003},
the present paper straightens the connection between the original technique and a true probabilistic model by
defining a coherent probabilistic model on the training dataset. This new model \eqref{eq:knn} then provides a
sound setting for Bayesian inference and for evaluating not just the most likely allocations for the
test dataset but also the uncertainty that goes with them. The advantages of using a probabilistic environment
are clearly demonstrated: it is only within this setting that tools like predictive maps as in Figure \ref{fig:levelset}
can be constructed. This obviously is a tremendous bonus for
the experimenter, since boundaries between most likely classes can thus be estimated and  regions can be established in which
allocation to a specific class or to any class is uncertain. In addition, the probabilistic 
framework allows for a natural and integrated analysis of the number of neighbours involved in the 
class allocation, in a standard model-choice perspective. This perspective can be
extended to the choice of the most significant components of the covariate $\bx$, even though this
possibility is not explored in the current paper. 

The present paper also addresses the computational difficulties related to this approach, namely the 
well-known issue of the intractable normalising constant. While this has been thoroughly discussed
in the literature, our comparison of three independent approximations leads to the strong conclusion 
that the pseudo-likelihood approximation is not to be trusted for training sets of moderate  size. Furthermore, 
while the path sampling and
perfect sampling approximations are useful in establishing this conclusion, they cannot be advocated at the
operational level, but we also demonstrate that a Gibbs sampling alternative to the perfect
sampling scheme of \cite{Moeller:Pettitt:Reeves:2006} is both operational and practical.

\section*{Acknowledgements}
The authors are grateful to Gilles Celeux for his numerous and 
insightful comments on the different perspectives offered by
this probabilistic reassessment, as well as to the Associate 
Editor and to both referees for their constructive comments. Both
JMM and CPR are also grateful to the Department of Statistics of the 
University of Glasgow for its warm welcome during various
visits related to this work.  This work was supported in part 
by the IST Programme of the European Community, under the
PASCAL Network of Excellence, ST-2002-506778.

\bibliographystyle{apalike}
\bibliography{CMRT07}

\end{document}